\documentclass[twocolumn,showpacs,nofootinbib,prd,amssymb,floatfix]{revtex4-2}
\usepackage{latexsym}
\usepackage{hyperref}
\usepackage{amsmath}
\usepackage{relsize}
\usepackage{dsfont}
\usepackage{todonotes}
\usepackage{textcomp}
\usepackage{float}
\newcommand{\aap}{Astron.\ Astrophys.}
\newcommand{\mnras}{Mon.\ Not.\ R.\ Astron.\ Soc.}

\newcommand{\apjl}{Astrophys.\ J.\ Lett.}

\newcommand{\araa}{Ann.\ Rev.\ Astron.\ Astrophys.}
\newcommand{\nphysa}{Nucl.\ Phys.\ A}

\newcommand{\del}[1]{}

\begin{document}

%%%%%%%%%%%%%%%%%%%%%%%%%%%%%%%%%%%%%%%%%%%%%%%%%%%%%%%%%%%%%%%%%%%%%%%%%%%%%%
\title{Thermodynamically consistent accreted crust of neutron stars: 
\\The role of proton shell effects}
%%%%%%%%%%%%%%%%%%%%%%%%%%%%%%%%%%%%%%%%%%%%%%%%%%%%%%%%%%%%%%%%%%%%%%%%%%%%%%

%%%%%%%%%%%%%%%%%%%%%%%%%%%%%%%%%%%%%%%%%%%%%%%%%%%%%%%%%%%%%%%%%%%%%%%%%%%
\author{Mikhail E.\ Gusakov}
%\thanks{}
\affiliation{Ioffe Institute, Saint-Petersburg, Russia}

\author{Andrey I.\ Chugunov}
%\thanks{}
\affiliation{Ioffe Institute, Saint-Petersburg, Russia}
%%%%%%%%%%%%%%%%%%%%%%%%%%%%%%%%%%%%%%%%%%%%%%%%%%%%%%%%%%%%%%%%%%%%%%%%%%%

\date{\today}

%%%%%%%%%%%%%%%%%%%%%%%%%%%%%%%%%%%%%%%%%%
\begin{abstract} 
Observations of accreting neutron stars are widely used to constrain the microphysical properties of superdense matter.
A key ingredient in this analysis is the heating associated with nuclear reactions in the outer layers of the neutron star (crust), as well as the equation of state and composition of these layers.
As recently shown, the neutron hydrostatic/diffusion (nHD) condition is valid in the inner part of the crust, where some of the neutrons are not bound to the nuclei, and this condition should be properly incorporated into crustal models. Here we construct  models of the
accreted crust of a neutron star, taking into account the nHD condition and 
proton shell effects in nuclei.
For numerical illustration, we employ the recently proposed compressible liquid drop model, which incorporates shell effects. However, our approach is general and can also be used in future studies relying on more sophisticated nuclear physics models.
\end{abstract}
%%%%%%%%%%%%%%%%%%%%%%%%%%%%%%%%%%%%%%%%%%
\date{\today}

%\pacs{	}

%%%%%%%%%%%%%%%%%%%%%%%%%%%%%%%%%%%%%%

\maketitle

\section{Introduction}\label{Sec_introd}

The observational properties of accreting neutron stars evolve on the timescale of a human 
lifetime, offering an opportunity to explore the neutron star ``real-time'' dynamics. Specifically, 
crust cooling following an accretion episode has been observed and analyzed for nine sources, 
\cite{syhp07,bc09,pr12,wdp17,Brown_ea18,mdkse18,pc21,Lu_ea22,Mendes_ea22,Page_ea22_Hyperburst,pgc23},
 while a few dozens of other accreting neutron stars in quiescence
demonstrate thermal emission from the fully thermally relaxed crusts \cite{bbr98,ylh03,ylpgc04,yp04,hjwt07,heinke_et_al_10,wdp13,by15,hs17,pcc19, 
Fortin_ea21,Mendes_ea22,Jain_ea23,pgc23}.
The necessary ingredients for modelling 
these sources are the crustal equation of state (EOS) 
and the heat release profile over the crust. 
Here, we determine these properties taking into account
the proton shell effects in nuclei
as well as the presence of unbound neutrons 
in the inner crust.

To construct the EOS of the accreted crust, one should study the accretion-driven evolution of volume elements in the crust. Namely, accretion leads to the compression of each volume element, initiating nuclear reactions there.
In early works, beginning from \cite{Sato79}, this problem was considered in a single-fluid approximation, i.e., it was assumed that all matter is 
confined within the compressing volume element,
thereby making pressure the only driving parameter of nuclear evolution
(the temperature effects were also neglected).

However, as indicated in \cite{GC20_DiffEq}, the problem is not that simple due to the presence of unbound neutrons in the inner crust. The unbound neutrons must be treated as an independent fluid, which makes the traditional (single-fluid) approximation inapplicable (see \cite{CS20_NoEquil} for a discussion of inconsistencies arising in the single-fluid approximation).
The behavior of the neutron fluid is quite simple: owing to 
superfluid motions
or rapid diffusion, it redistributes itself within the inner crust to 
remain in the 
hydrostatic/diffusion equilibrium (nHD) 
 governed by the 
{\it nHD condition}
\begin{equation}
\mu_n^\infty={\rm constant},
\label{nHD}
\end{equation}
where
$\mu_n^\infty=\mu_n e^{\nu/2}$ is the redshifted neutron chemical potential, $\mu_n$ is
the local neutron chemical potential,
and $e^{\nu/2}$  is
the redshift factor.
By definition, inner crust 
corresponds to $\mu_n\ge m_n c^2$ ($m_n$ is the bare neutron mass; $c$ is the speed of light) and the upper boundary of the inner crust is given by the condition 
$\mu_n=m_n c^2$. 
In the approximation of vanishing stellar temperature, $T=0$,
there are no unbound neutrons above the inner crust. 
The continuity of the neutron chemical potential at the crust-core boundary, as well as the equilibrium structure of the core, 
implies that the redshifted neutron chemical potential 
is constant both in the inner crust and core \cite{GC20_DiffEq}.

Because of neutron leakage, it is essential to associate a volume element with the nuclei and monitor nuclear reactions 
in that element.
The nuclear evolution is governed by two key parameters: the pressure determined by the hydrostatic 
crustal model above this volume element and the neutron chemical potential, 
which can also be affected by the crustal model below the chosen volume element due to the nHD condition.
As a result, it is generally not possible to build a model of the inner crust layer by layer, starting from the top; instead, it is necessary to consider the nuclear evolution in the entire crust simultaneously.

This problem is especially complicated (and has not been analyzed yet) at the initial stages of accretion, when the original composition of the matter in the pristine crust 
is replaced by the accreted material. 
We will not consider this transitional regime in what follows. 
Instead, as in our previous works, we will focus on investigating the steady-state regime of accretion, in which the composition of the crust no longer depends on time (except for small secular corrections associated with changes in the mass 
of the accreting star). 
The resulting neutron star crust will be referred to as the fully accreted crust (FAC).

The assumption of FAC makes the problem self-similar and substantially simplifies it.
In particular, if $P_\mathrm{oi}$, the pressure at the outer-inner crust interface (oi), is known, it becomes possible to construct 
a model of the crust, starting from the top of the crust 
and considering it layer by layer.
The equations governing the nHD crust were first derived in \cite{GC20_DiffEq} and rederived here 
in Section \ref{Sec_Eqs} for a more realistic nuclear physics model that includes shell effects.

Unfortunately,
generally $P_\mathrm{oi}$ cannot be known in advance and must be determined as a result of FAC 
modeling. To do this, we treat $P_\mathrm{oi}$ as a parameter and apply the equations from Section 
\ref{Sec_Eqs} to construct the nHD EOS family, parametrized by the pressure $P_\mathrm{oi}$.
We then analyze this family to constrain the actual value of $P_\mathrm{oi}$ and the corresponding FAC model according to two requirements.

First, the number of nuclei in the crust must be nearly constant for the crustal structure to remain self-similar. 
Because accretion supplies additional nuclei to the crust, there must be an effective mechanism for nuclei disintegration. 
The physical mechanism 
for disintegration is related to a specific 
instability, which was identified in \cite{GC20_DiffEq} and further analyzed here (see Section \ref{Sec_Res}). According to the numerical results, this instability occurs if $P_\mathrm{oi}$ exceeds the critical value $P^\mathrm{(min)}_\mathrm{oi}$, thus constraining $P_\mathrm{oi}$ from below.

The second condition is used to determine the upper bound for $P_\mathrm{oi}$. It arises from the requirement that the FAC must be thermodynamically consistent with the neutron star core.
It is important to note that the self-similar solution ends at the point where all the nuclei 
disintegrate and generally cannot be continued into the underlying layers.
For the compressible liquid drop  (CLD) model used in \cite{GC20_DiffEq}, 
this is not a problem, as the FAC solution ends at the crust-core boundary.
However, for more realistic models, the
situation is not that simple and relic crustal layers may 
remain between the FAC and the core (see Ref.\ \cite{GC21_HeatReleaze} and Sections \ref{Sec_Res} and \ref{Sec_Poi}).

For numerical illustration (Section \ref{Sec_Res}) we limit ourselves to a pure $^{56}$Fe ash composition (see Refs.\ \cite{SGC_OC21,SGC22,SGC23_compos} for multicomponent models, which, however, are limited to not-too-deep crustal layers) and apply the recently suggested CLD model with proton shell effects added on top (CLD+sh model, \cite{carreau_ea20_Cryst_CLDsh}).
Using this model, we 
constrain the pressure $P_\mathrm{oi}$ in Section \ref{Sec_Poi}.
In Section \ref{Sec_Qdeep}, we analyze heat release at the innermost regions of inner crust and, in Section \ref{Sec_Heurist}, present a heuristic energy-based approach to predict FAC properties for more refined nuclear physics models.
Our conclusions and results
are summarized in Section \ref{Sec_Summary}.

%%%%%%%%%%%%%%%%%%%%%%%%%%%%%%%%%%%%%%%%%%%%%%%%
\section{Construction of nHD crust} \label{Sec_Eqs}
%%%%%%%%%%%%%%%%%%%%%%%%%%%%%%%%%%%%%%%%%%%%%%%%
As an input for the development of
accreted crust models, it is necessary to invoke two fundamental physical theories: thermodynamics and kinetics of crustal matter.
Thermodynamics is required to calculate the equation of state,
assuming a given chemical composition and local thermodynamic equilibrium. The equation of state, determined in this manner, will be
referred to as the microscopic equation of state  (mEOS); note that we distinguish it from the actual equation of state established in a specific accreting neutron star.
In turn, kinetics is 
necessary to consistently determine the composition of matter in each point of the accreting crust by accounting for nuclear reactions and the redistribution of free neutrons within the crust.

In Ref.\  \cite{GC20_DiffEq}, we analyzed the nHD crust using the smooth CLD model as our 
thermodynamic 
framework.
The corresponding mEOS is two-parametric, i.e., the energy density $\epsilon=\epsilon(n_{b},n_N)$ is a function of the baryon number density $n_{b}$ and the number density of nuclei, $n_N$. In Section \ref{Sec_EqsSmooth} we rederive the equations of Ref.\ \cite{GC20_DiffEq} in a more general form, which simplifies the subsequent discussion.

The main goal of this paper is to consider the nHD crust with a more realistic microphysics input that includes shell effects.
This significantly complicates the problem because of two reasons. 
Firstly, mEOS ceases to be two-parametric
(see Section \ref{Sec_Micr_Shell} for
details). 
As a result, the equations governing the nHD crust need to be modified to be consistent with this more realistic mEOS. The corresponding modification is presented in Section \ref{Sec_EqsShell}.
The algorithm for the construction of the nHD crust based on these equations is given in Section \ref{Sec_Algorithm}.

Secondly, the calculation of the shell effects to determine mEOS is a complicated and model-dependent problem. In this work, we apply three simplified models to describe shell energy corrections (see Section \ref{Sec_Shell_Models}). 
We treat these models more as qualitative ones,
and use them
as a proof-of-principle 
demonstration of 
how one should proceed in order to construct the nHD accreted crust accounting for shell effects.
We find that the latter effects are crucial for modelling the nHD crust and infer
some general trends that appear to be 
less sensitive to the quantitative behavior of shell corrections.

\subsection{Smooth mEOS}  \label{Sec_EqsSmooth}

In Ref.\ \cite{GC20_DiffEq} we started the derivation of the equations for the nHD crust from the two-parametric mEOS $\epsilon(n_{b},n_N)$ and used the CLD model for numerical illustration. The CLD-based mEOS can be represented in this form by 
imposing beta-equilibrium, mechanical equilibrium, and chemical equilibrium conditions inside a spherical cell, which contains one nucleus (see the Supplementary Material of that work). Here, we take a step back and
start from a three-parameter mEOS, $\epsilon(n_{b},n_N, Z)$ (see Section \ref{Sec_Micr_Smooth} for numerical implementation). This allows us to consider not only beta-equilibrated matter, but also a EOS for which $Z$ is constant 
in the inner crust. 
The latter model, referred to as ``$Z$-fixed'' EOS in what follows, is used as a simplified, yet 
adequate model mimicking the strong proton shell closure of nuclei with $Z=20$.

Let us introduce the chemical potentials, $\mu_N=\partial \epsilon/\partial n_{N}$ and $\mu_{b}=\partial \epsilon/\partial n_{b}$.
The first one, $\mu_N$, describes the energy change resulting from the addition of a nucleus at fixed $n_{b}$ and $Z$ (alternatively, one can consider it as the creation of a nucleus from nucleons already available in the matter),
while $\mu_n$ corresponds to the energy change due to an additional baryon at fixed $n_N$ and $Z$.
Since the proton number density, $Z n_N$, remains also unchanged, the added baryon is a neutron and $\mu_b$ can be identified with the neutron chemical potential, $\mu_n$: 
$\mu_b=\mu_n$. 
Below, we will write $\mu_n$ instead of $\mu_b$ in all formulas.

In this section, we consider the two cases:  (i) $Z$ is conserved during compression ($Z$-fixed EOS), and (ii) beta-equilibrated matter, for which (see Section \ref{Sec_Micr_Smooth})
\begin{equation}
    \left. \frac{\partial \epsilon}{\partial Z}\right|_{n_b,\, n_N}=0.
\label{betaEq_Smooth}
\end{equation}
In both cases, the pressure can be written as
\begin{equation}
    P=-\frac{\partial (\epsilon V)}{\partial V}=-\epsilon + \mu_{n} n_{b}
     +\mu_N  n_N. 
\label{P_smooth}
\end{equation}
Here, the second equality can be  derived straightforwardly by taking partial derivative at a fixed nucleon and baryon number, and using the definitions of the chemical potentials $\mu_{n}$ and $\mu_{N}$; see also Section \ref{Sec_EqsSmooth}.

According to one of the Tolman-Oppenheimer-Volkoff equations
\cite{hpy07}
\begin{equation}
    P^\prime=-(P+\epsilon)\nu^\prime/2.
    \label{TOV}
\end{equation}
Here the prime denotes the derivative with respect to the radial coordinate $r$. Combined with the nHD condition (\ref{nHD}) and the Gibbs-Duhem relation in the form $dP=n_{b} d\mu_n +n_N d\mu_N$, which holds true for both the $Z$-fixed EOS and beta-equilibrated EOS, we arrive at the condition $\mu_N^\infty=\mathrm{constant}$.
Thus,
\begin{equation}
\mu_N=C \mu_n.
\label{cat3}
\end{equation}
Here $C$ is a constant that depends on the pressure $P_{\rm oi}$ at the outer-inner crust boundary. 
As we demonstrated in Ref.\ \cite{GC20_DiffEq} (see also Section \ref{Sec_Poi}),
FAC EOS corresponds to a certain value of $C$.
However, to determine this value, it is instructive to consider the whole nHD EOS family, i.e., a family of EOSs that are allowed by the nHD condition (\ref{nHD}). The nHD EOS family can parametrized by the pressure $P_{\rm oi}$ (or, equally, by $C$). 

The catalyzed crust corresponds to the global minimum of the energy density 
$\epsilon(n_b, n_N, Z)$ at fixed $n_b$, which is given by the beta-equilibrium condition (\ref{betaEq_Smooth}) and the condition 
$\mu_N= \partial \epsilon/\partial n_N=0$. 
As pointed out in Ref.\ \cite{GC20_DiffEq}, it is a member of the beta-equilibrium nHD EOS family, corresponding to $C=0$.

\subsection{Realistic mEOS }  \label{Sec_EqsShell}

For a realistic modeling of the accreted crust, one should utilize the mEOS, which takes into account nuclear shell effects. 
Obviously, the corresponding mEOS is more complicated than its CLD-based smooth
analogue. 
In particular, the nuclear charge  number $Z$ becomes discrete (integer), and the energy density dependence on $Z$ becomes rather 
complicated
(see, e.g., Refs.\ \cite{Pearson_ea18_bsk22-26,carreau_ea20_Cryst_CLDsh} and Section \ref{Sec_Micr_Shell}).
As a result, the beta-equilibrium condition (\ref{betaEq_Smooth})
cannot be applied and should generally be replaced with some other requirement.
Moreover, Eq.\ (\ref{cat3}) can be violated if $Z$ varies 
in the inner crust due to nuclear reactions. 
To address these difficulties, we should modify our approach as discussed in Ref.\ \cite{GKC_psi21}.

Namely, we should replace Eq.\ (\ref{cat3}) with its more general counterpart
which, similarly to Eq.\ (\ref{cat3}) follows from
the Tolman-Oppenheimer-Volkoff equation (\ref{TOV})
and nHD condition (\ref{nHD}):
\begin{equation}
\mu_{n}=m_{n} c^2  \, 
    \exp
        \left[\int_{P_{\rm oi}}^P {\rm d}\widetilde{P}/(
         \epsilon(\widetilde{P})
         +\widetilde{P})\right],
\label{munn2}
\end{equation}
where, as we already indicated in the introduction, $m_n c^2$ is the neutron chemical potential at the top of the inner crust (located at $P=P_{\rm oi}$). This equation allows us to determine  the chemical potential $\mu_n$ in the layer with the pressure $P$, if the EOS [e.g., the function $\epsilon(P)$] is known in the range from $P_{\rm oi}$ to $P$.
In accordance with Refs.\ \cite{HZ90, Steiner12, SC19_MNRAS}, 
our approach to constructing the inner crust model involves a step-by-step progression into deeper layers of the crust. At each layer, we determine the charge number $Z$ by minimizing the corresponding thermodynamic potential 
by means of
permissible nuclear reactions 
(see Section \ref{Sec_Algorithm} for more details). 
Previously (e.g., in Refs.\ \cite{HZ90,Steiner12,SC19_MNRAS}), the redistribution of unbound neutrons was disregarded. 
Consequently, the reactions were assumed to occur at a constant pressure, and the potential to be minimized was associated with the Gibbs free energy
(e.g., \cite{HZ90,Steiner12}). 
However, as stipulated by Eq.\ (\ref{munn2}), 
in the nHD crust, not only is the pressure $P$ fixed 
in a given layer, but also the chemical potential $\mu_n$. As shown in \cite{GKC_psi21}, the appropriate thermodynamic potential that should be minimized at fixed $P$ and $\mu_n$ is then 
\begin{equation}
    \Psi=(\epsilon+P)V-\mu_n N_{b}.
    \label{Psi}
\end{equation}
Here $V$ is the volume attached to nuclei, and 
$N_{b}=V n_b$ is the total number of baryons in the volume.

If (pycnonuclear) fusion and fission reactions are not allowed (either too slow or energetically forbidden), the number of nuclei is conserved. In this case, to determine $Z$ one can minimize the potential $\Psi$ per one nucleus (note that it coincides with $\mu_N$ 
in a special case considered in this work when nuclei of only one species are present at any given pressure in the crust):
\begin{equation}
    \psi\equiv\frac{\Psi}{V n_N}=\frac{\epsilon+P-\mu_n n_{b}}{n_N}=\mu_N.
    \label{psi3}
\end{equation}
This minimization 
effectively replaces the 
beta-equilibrium condition. For a smooth CLD model with continuous $Z$, it  reduces to the equation 
(\ref{betaEq_Smooth}); see Section \ref{Sec_Micr_Shell} for details.

When the composition of a given layer is determined, we can apply Eq.\ (\ref{munn2}) to consider the underlying layer. By repeating this procedure, we develop a step-by-step algorithm, which is formulated in the next subsection and applied in this work to construct the nHD crust.

\subsection{Algorithm }  \label{Sec_Algorithm}

Similar to the case of a smooth mEOS, we begin by constructing the nHD EOS family, which is 
parametrized by the pressure $P_\mathrm{oi}$ at the outer-inner crust interface.
In subsequent sections, we discuss how to constrain the range of realistic $P_\mathrm{oi}$ 
and present numerical results for the CLD+sh model of Ref.\ \cite{carreau_ea20_Cryst_CLDsh}.

To build the nHD EOS family, we apply an algorithm based on Eqs.\ (\ref{munn2}) and (\ref{Psi}).
We stress that this algorithm is quite general and, in particular, 
applicable in the situation when both shell effects and odd-even staggering of nuclear energies are 
allowed for.
Since the thermodynamic potential $\Psi$ should be minimized at fixed $P$ and $\mu_n$ we assume the mEOS to be parameterized by the pressure $P$, neutron chemical potential $\mu_n$, and 
nuclear charge number $Z$. In other words, the microphysical model should allow one to calculate 
the energy density $\epsilon$ and potential $\psi$  for a given $P$, $\mu_n$, and $Z$ (see Section 
\ref{Sec_Micr_Shell} for the realization of this form of mEOS based on the CLD+sh model, which is 
used in this work
as a numerical example).

The algorithm contains the following stages (stages 2 and 3 are repeated at each step):
\begin{enumerate}
	
	\item {\it Specify  the initial conditions at the top of the inner crust.}
 
    The EOS of the fully accreted outer crust is well studied, especially for a one-component ash 
    discussed here (see, e.g., Refs.\ \cite{HZ90,cfzh20}). This allows us to determine the charge 
    number at the bottom of the outer crust, located at $P=P_\mathrm{oi}$. 
    Combining with the neutron chemical potential at the top of the inner crust, which is equal to  
    $\mu_{n,{\rm oi}}=m_n c^2$ by definition, we obtain the initial conditions (the $j=0$ step) for 
    the construction of the nHD inner crust. Before turning to the next stage, we also analyze the 
    reaction pathways at the oi interface according to stage 3 to determine the composition at the 
    top of the inner crust.
    
	\item {\it Advancing to $j$-th layer.}
 
	Starting this step, we assume that the equation of state in the previous, $(j-1)$-th layer, has been determined and the respective pressure is $P_{j-1}$, the neutron chemical potential 
    is $\mu_{n,j-1}$, and the charge number is $Z_{j-1}$.
    We increase the pressure by a small amount $\Delta P$, 
    such that $P_{j}=P_{j-1}+\Delta P$. To guarantee the nHD equilibrium, we also increase the neutron chemical potential  
    according to the formula  (\ref{munn2}):
    \begin{align}
    \mu_{n,j}=\mu_{n, j-1} \, \cdot \exp\left[ \frac{\Delta P}{
            \epsilon(P_{j-1})+P_{j-1}}\right]
    \label{munj}
    \end{align}
    The composition will be adjusted to the 
    equilibrium one at the next stage, and here we simply assign $Z_{j}=Z_{j-1}$.
	
	\item
     {\it Choosing optimal $Z_j$}
	
	As inputs for this stage, we have the pressure $P_{j}$ and neutron chemical potential $\mu_{n,j}$ at the layer $j$, but the nuclear
    charge $Z_{j}$ may differ from the optimal value. 
    It can either 
    be equal to $Z_{j-1}$ 
    or correspond to the value obtained in
    the preceding iteration of this stage. 
    In order to check whether $Z_j$ is further changing,
    we calculate the potential $\psi$  for $Z=Z_{j}$, $Z_{j}-1$, and $Z_{j}+1$. The latter two are nuclei that can be produced by electron capture and emission, respectively.
	By comparing the values
	$\psi(Z_{j})$, $\psi(Z_{j}-1)$, and $\psi(Z_{j}+1)$ we choose 
	$Z$ corresponding to the minimal $\psi$
    and assign $Z_j = Z$.
	If $Z_j$ is modified, we repeat this stage
    again.
	If it is not modified, $Z_j$  
    is considered optimal and we 
	proceed to the  next crustal layer by applying the algorithm of stage 2.
		
\end{enumerate}

In principle, the fourth stage, which checks for pycnonuclear reactions 
at the layer $j$, can be added. 
However, for $^{56}$Fe ash this stage is 
unnecessary,
because 
the typical $Z\approx 20$, which is realized in the inner crust in this case,
is too large, leading to extremely small rate of pycnonuclear reactions.

Obviously, this algorithm should be 
interrupted
if the crust-core boundary is reached, i.e., 
when, for some $\mu_n$, the pressure at the crust 
matches that in the core.

In fact, 
the algorithm can 
be interrupted even earlier because of
the disintegration of all nuclei at stage 3, before reaching the crust-core boundary.
This disintegration 
could take place
as a result of the onset of an instability at $P=P_{\rm inst}$, as described in Ref.\ \cite{GC20_DiffEq} for the smooth CLD model.
This is the most interesting case, because the instability is required for the formation of FAC.

%%%%%%%%%%%%%%%%%%%%%%%%%%%%%%%%%%%%%%%
\begin{figure*}
	\includegraphics[width=0.992\textwidth]{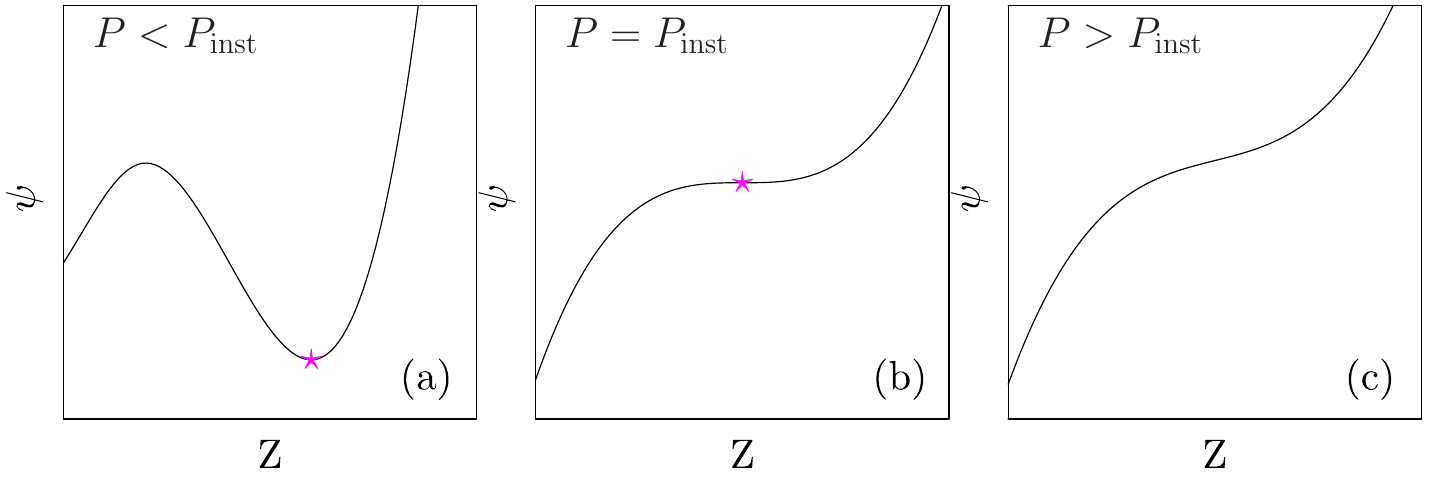}
	\caption{Qualitative behaviour of the potential $\psi$ 
        for a smooth model with continuous $Z$
        for the three values of pressure: $P<P_\mathrm{inst}$, $P=P_\mathrm{inst}$, and $P>P_\mathrm{inst}$.}
	\label{Fig_psi1}
\end{figure*}
%%%%%%%%%%%%%%%%%%%%%%%%%%%%%%%%%%%%%%%

In terms of the potential $\psi$ the criteria for the instability can be expressed as
$\psi(Z_{\rm inst})=\psi(Z_{\rm inst}-1)>\psi(Z_{\rm inst}-2)> \ldots >\psi(Z=1)>0$,
where  $Z_{\rm inst}$ is the charge at the previous step of the algorithm (at the layer $P<P_\mathrm{inst}$). In principle, pycnonuclear fusion can become important once $Z$ has decreased to a low enough value during the disintegration process. 
In this case, further transformation of nuclei into neutrons 
proceeds via an unstoppable analogue of the superthreshold electron capture cascade (SEC, \cite{gkm08}). This modification does not affect 
the final outcome -- all nuclei disintegrate  in the considered layer.
Specifically, fusion-produced nuclei will undergo beta captures and subsequent neutron emissions, 
leading to low-$Z$ nuclei, which, in turn, undergo pycnonuclear fusion. 
An overall result of this process is disintegration of one nucleus $(Z,A)$ per one pycnonuclear reaction and production of $A$ neutrons, which redistribute over the crust 
to keep fixed $\mu_n$ in the instability layer:
$(Z,A)+(Z,A)\rightarrow (2Z, 2A) \rightarrow (Z,A) + A n$ ($A$ is the atomic mass number).
It should be noted that in the 
traditional
approach that neglects neutron redistribution, the SEC cascade 
increases the number of
unbound neutrons, affecting
$\mu_n$ 
and preventing complete disintegration in some layers
(e.g., \cite{lau_ea18,SC19_MNRAS}). 
In contrast,
for the nHD crust, the SEC cascade becomes unstoppable: the nHD condition fixes the neutron 
chemical potential and thus the amount of free neutrons in the layer. The neutrons produced by 
disintegration of nuclei are removed from the layer by superfluid flow or diffusion.

To illustrate the onset of the instability, let us consider  the ``smooth'' model,  
i.e., by treating $Z$ as a continuous variable.
The schematic $\psi(Z)$ profiles for 
such model are shown 
in Figure \ref{Fig_psi1}.
Panel (a) shows the typical profile for layers located at $P<P_{\rm inst}$. The function $\psi(Z)$ has a profound minimum, shown by the star, where nuclei are stable.
However, there is also another extremum (local maximum) at a lower $Z$.
With
pressure increase,
the minimum and the maximum become closer and closer, and at $P=P_\mathrm{inst}$, they 
merge,
producing an inflection point, shown by the star in panel (b).
At this point, nuclei become unstable and undergo a series of beta captures ($Z$ is lowered),
until
complete disintegration into neutrons.
Clearly, 
 at the instability point, we have
\begin{eqnarray}
    0&=&\left.\frac{\partial \psi}{\partial Z}\right|_{P_{\rm inst},\,\mu_{n,{\rm inst}}},
        \label{inst1}\\
    0&=&\left.\frac{\partial^2 \psi}{\partial Z^2}\right|_{P_{\rm inst},\,\mu_{n,{\rm inst}}}.
        \label{inst2}
\end{eqnarray}
At $P>P_\mathrm{inst}$ (panel c)  $\psi$ monotonically increases with 
$Z$, 
hence no beta-stable solutions [see Eq.\ (\ref{inst1})] are available.

In a more realistic model,
the shell effects introduce 
many local minima in the $\psi(Z)$ curve (see Figure \ref{Fig_PsiProf_shell}).
However, for sufficiently large $P$, the general slope of the function $\psi(Z)$ becomes strong enough to 
smooth out
the local minima, leading to the instability.

As long as all nuclei in the considered volume disintegrate, the construction of the nHD accreted 
crust cannot be continued unambiguously to higher 
pressures.
However, reaching $P=P_\mathrm{inst}$ does not necessarily mark the end of the crust. 
At larger pressures, $P>P_\mathrm{inst}$, the crust
can be continued by ``relic'' layers that
are not replaced
by the accreted material in the FAC regime.
These layers are formed at the initial stages of accretion and can be composed of the spherical nuclei or more complicated nuclear shapes referred to as ``pasta'' (see, e.g., \cite{Fattoyev+17,Vinas+17,ch_17,Schneider_ea18,Pearson+20,JiHuShen21,DinhThi_ea21_CLD_pasta,pc22_pasta,Newton+22,scp23} for a recent discussion of pasta in pristine crust). 
Clearly, the unambiguous determination of the composition of these layers requires consideration of their formation, which is beyond the scope of the present paper.

\section{Microphysics input} \label{Sec_Micr}

In our previous work \cite{GC20_DiffEq}, we employed the CLD model based on the extended Thomas-Fermi (ETF) calculations of the nucleus surface properties,
which explicitly incorporates the neutron skin effects
(see detailed description  in the Supplementary Materials of Ref.\ \cite{GC20_DiffEq}).

In this paper, we apply the CLD and CLD+sh models  of Ref.\ \cite{carreau_ea20_Cryst_CLDsh}, which are slightly different. The basic smooth CLD model 
of that reference
does not explicitly take into account the neutron skin effect, and the surface properties are fitted to reproduce ETF calculations of the 
nuclear mass table
for the HFB24 model 
(see \cite{carreau_ea20_Cryst_CLDsh} for details).
As a result of this difference, the set of variables for the smooth CLD model of Ref.\ \cite{carreau_ea20_Cryst_CLDsh} differs from \cite{GC20_DiffEq}.
This requires
some additional derivations in order to rewrite 
the CLD model in terms of the variables that are useful for constructing the nHD crust (Sections 
\ref{Sec_EqsSmooth} and \ref{Sec_EqsShell}).
These derivations are rather straightforward, but we present them in the Section \ref{Sec_Micr_Smooth} for completeness.

The CLD+sh model of Ref.\ \cite{carreau_ea20_Cryst_CLDsh} is more realistic than the smooth CLD model of Ref.\ \cite{GC20_DiffEq},
because it includes proton shell effects (neutron shell corrections are small
and can be 
disregarded
\cite{chamel06,cnkm07}).
The proton shell energies are added on top of the CLD model. They are determined in Refs.\ \cite{Pearson_ea18_bsk22-26,Pearson_ea19_ShellCorr_Errata} from the ETF plus Strutinsky integral method (ETFSI).
As shown in \cite{carreau_ea20_Cryst_CLDsh}, the resulting CLD+sh model reproduces the most realistic  calculations of the inner crust 
to date 
\cite{Pearson_ea19_ShellCorr_Errata}, providing a unique tool to study the nHD crust.
The reduction of the CLD+sh model to the variables of Section \ref{Sec_EqsShell} is presented in section \ref{Sec_Micr_Shell}. 

\subsection{Smooth CLD model} \label{Sec_Micr_Smooth}

When applying the CLD model without shell effects, it is natural to assume that the nuclear charge $Z$ evolves  continuously in the inner crust.
Of course, these assumptions will be relaxed in the next section, where the shell effects 
are taken into consideration.
Here, we apply the CLD model suggested in Ref.\ \cite{carreau_ea20_Cryst_CLDsh}.

The CLD model of Ref.\ \cite{carreau_ea20_Cryst_CLDsh} starts from an explicit expression for the free energy  (see section 2 of that reference). In the zero temperature limit, which is 
adopted here,  it reduces to the energy density 
$\epsilon(n_{ni},n_{pi},n_{no},n_{N},n_e,w)$,
where $n_{ni}$ and $n_{pi}$ are, respectively, 
the neutron and proton number densities inside nuclei;
$n_{no}$ is the number density of unbound neutrons;
$n_{N}$ and $n_{e}$ are, respectively, the number densities of nuclei and electrons;
and $w=V_{p} n_{ N}$ is the fraction of volume occupied by nuclei
($V_p$ is the volume of a single nucleus).

Following Ref.\ \cite{GC20_DiffEq},
it is useful to introduce  the ``total''
number densities
$n_{ni}^{\rm (tot)}=n_{ni} w$,
$n_{pi}^{\rm (tot)}=n_{pi}w$, and $n_{no}^{\rm (tot)}=n_{no}(1-w)$, instead of  $n_{ni}$, $n_{pi}$, and $n_{no}$.
For example, $n_{ni}^{\rm (tot)}$ can be interpreted as the total number of neutrons in nuclei divided by the total  volume (not the volume occupied by nuclei).
Using these variables, 
the differential $ d \epsilon$ can be expressed as
\begin{eqnarray}
    d \epsilon&=& 
        \frac{\partial \epsilon }{\partial n_{n{ i}}^{\rm (tot)}} d n_{n{ i}}^{\rm (tot)}
        +\frac{\partial \epsilon }{\partial n_{p{ i}}^{\rm (tot)}} d n_{p{ i}}^{\rm (tot)}
        +\frac{\partial \epsilon }{\partial n_{n{ o}}^{\rm (tot)}} d n_{n{ o}}^{\rm (tot)}
\nonumber 
\\
    &+&\frac{\partial \epsilon }{\partial n_{ N}} d n_{ N}
      +\frac{\partial \epsilon }{\partial n_{ e}} d n_{ e}
      +\frac{\partial \epsilon }{\partial w} dw.
\label{dee2}
\end{eqnarray}
Introducing the baryon number density $n_{b}=n_{ni}^{\rm (tot)}+n_{pi}^{\rm (tot)}+n_{no}^{\rm (tot)}$,
the neutron chemical potentials inside $\mu_{ni}=\partial \epsilon/\partial n_{ni}^{\rm (tot)}$
and outside 
$\mu_{no}=\partial \epsilon/\partial n_{no}^{\rm (tot)}$ nuclei, 
the proton chemical potential inside nuclei, 
$\mu_{pi}= \partial \epsilon/\partial n_{pi}^{\rm (tot)}$,
as well as
the chemical potentials of nuclei
$\mu_N=\partial \epsilon/\partial n_{N}$,
and electrons,
$\mu_e=\partial \epsilon/\partial n_{e}$,
Eq.\ (\ref{dee2}) can be rewritten as
\begin{eqnarray}
    d \epsilon&=& 
        \mu_{ni} d n_{b}
        +(\mu_{pi}+\mu_e-\mu_{ni}) dn_{pi}^{\rm (tot)}
        +\mu_N d n_N 
\nonumber
\\
    &+&(\mu_{no}-\mu_{ni}) dn_{no}^{\rm (tot)}
    +\frac{\partial \epsilon }{\partial w} dw,
\label{dee3}
\end{eqnarray}
where we make use of the quasineutrality condition, $n_e=n_{pi}^{\rm (tot)}$.
It is 
worth noting, that $\mu_{pi}$ and $\mu_{ni}$ are not equal to their
bulk 
counterparts.
In particular, both these quantities include contributions from the surface energy, and $\mu_{pi}$ additionally includes a correction associated with Coulomb energy.

To obtain EOS in the form of Section \ref{Sec_EqsSmooth}, we should keep $n_{N}$, $n_{b}$, and $Z=n_{pi}^{\rm (tot)}/n_N$ fixed, while determining the remaining parameters  $w$ and $n_{no}^{\rm (tot)}$ 
by minimizing $\epsilon$
\begin{eqnarray}
    0&=&\frac{\partial \epsilon(n_{b},n_{pi}^{\rm (tot)},n_{no}^{\rm (tot)},n_N,w) }{\partial w},
  \label{difeq1}\\
    0&=&\frac{\partial 
    \epsilon(n_{b},n_{pi}^{\rm (tot)},n_{no}^{\rm (tot)},n_N,w)}{\partial n_{no}^{\rm (tot)}}=\mu_{no}-\mu_{ni},
  \label{difeq2}
\end{eqnarray}
These equations have a natural physical meaning: the first one represents the mechanical equilibrium of nucleus and unbound neutrons, 
while the second one represents the diffusion equilibrium for neutrons 
outside and inside nuclei
(an analogue of the equilibrium with respect to neutron emission and captures).
Since $\mu_{no}=\mu_{ni}$, below we refer to both of these quantities as the neutron chemical potential and denote it simply as $\mu_n$.

Let us consider the beta equilibrium condition (\ref{betaEq_Smooth}).
It is equivalent to $\partial \epsilon/\partial n_{pi}^{\rm (tot)}=0$, leading to
\begin{eqnarray}
    0&=&\mu_{pi}+\mu_e -\mu_n.
    \label{cat1}
\end{eqnarray}

The equilibrium (catalyzed) crust can be obtained by imposing an additional condition,
\begin{equation}
    \mu_N=\frac{\partial 
    \epsilon(n_{b},n_{pi}^{\rm (tot)},n_{no}^{\rm (tot)},n_N,w)}{\partial n_{N}}=0.
\label{cat2}
\end{equation}
This condition 
allows one to find
the optimal (equilibrium) number density of nuclei,  $n_N$.

For the accreted crust, which is considered here, additional nuclei are provided by accretion, leading to a nonequilibrium  $n_N$ and nonequilibrium crust \cite{GC20_DiffEq,GC21_HeatReleaze}.
As shown in Section \ref{Sec_EqsSmooth}, for the nHD inner crust, Eq.\
(\ref{cat2}) should be replaced by  Eq.\ (\ref{cat3}).

For the sake of completeness, let us point out that
the minimization of the $\psi$ potential with respect to $Z$ can be shown to be equivalent to the beta-equilibrium condition 
\begin{equation}
    \left.\frac{\partial \psi}{\partial Z}\right |_{P, \, \mu_n}=\mu_{pi}+\mu_e -\mu_n=0.
\label{cldm}
\end{equation}

Summarizing, the equations (\ref{difeq1})--(\ref{cat1}) and (\ref{cat3}) constitute a complete system of equations that allows one to calculate all the thermodynamic quantities if the parameter $C \equiv \mu_N/\mu_n$
and one of the thermodynamic quantities (e.g., $n_{b}$ or pressure $P$) are given.

\subsection{CLD+sh model} \label{Sec_Micr_Shell}

Following Ref.\ \cite{carreau_ea20_Cryst_CLDsh}, we derive the CLD+sh model by constraining the nuclear charge $Z$ in the smooth CLD model to integer values and incorporating
precalculated shell energies.
Consequently, the total energy density is expressed in the form
\begin{equation}
    \epsilon(n_{b}, n_N, Z, n_{no}^\mathrm{(tot)}, w)=\epsilon^{\rm CLD}+\Delta \epsilon^{\rm shell},
\label{e1}
\end{equation}
where we employ the notations of Section \ref{Sec_Micr_Smooth}; 
$\epsilon^{\rm CLD}$ is the energy density as 
it is 
given by the 
CLD model, and $\Delta \epsilon^{\rm shell}$ is a correction to the energy density, associated with the shell effects.

It is  important to emphasize  that in Refs.\ \cite{Pearson_ea12,Pearson_ea15_pairing,Pearson_ea18_bsk22-26,carreau_ea20_Cryst_CLDsh} the shell corrections were primarily applied  to determine  $Z$ for the ground state composition, i.e., $Z$ that minimizes energy density at a fixed $n_b$.
However, shell corrections appear to have been overlooked when calculating the ``secondary'' thermodynamic quantities, such as pressure. As a result, the pressure, as it was calculated in \cite{Pearson_ea12,Pearson_ea19_ShellCorr_Errata}, may not be fully thermodynamically consistent.
While this inconsistency is likely insignificant 
for determining the
catalyzed EOS, it could be important in the context of the energy release in the accreted crust discussed here.
Therefore, we endeavor to avoid this inconsistency.

In general,  
proceeding within the CLD
approach, 
the shell correction $\Delta \epsilon^{\rm shell}$  to the energy density 
should be considered as a function of 
five independent thermodynamic variables: 
$n_{b}$, $n_N$, $Z$, $n^\mathrm{(tot)}_{ni}$, and $w$.
However, in this section, for the sake of simplicity,  we postulate that the shell energy corrections can be expressed in the form
\begin{equation}
    \Delta \epsilon^{\rm shell}=n_{N} E^{\rm shell}(n_{b},n_{N},Z).
  \label{shell_CLD_gen}
\end{equation}
This assumption is made taking into account that the shell corrections are computed on top of the CLD model, which is minimized over internal CLD variables ($n^\mathrm{(tot)}_{no}$ and $w$) according to Eqs.\ (\ref{difeq1}) and (\ref{difeq2}).
As a result, we arrive at the three-parameter mEOS, 
with the energy density
$\epsilon(n_{b}, n_N, Z)$, which should be used to  calculate the 
remaining
thermodynamic quantities consistently. In particular, the pressure can be determined as
\begin{equation}
    P=-\frac{\partial (\epsilon V) }{\partial V}=
     -\epsilon + \mu_{b} n_{b}
     +\mu_N  n_N,
   \label{P_shell}
\end{equation}
where the chemical potentials 
are given by
\begin{eqnarray}
    \mu_b&\equiv&\frac{\partial\epsilon(n_{b}, n_N, Z)}{\partial n_b}
    =\mu_n 
    =\mu_n^\mathrm{CLD}+\mu_n^\mathrm{shell}, \label{mu_n_shell}
    \\
    \mu_N&\equiv&\frac{\partial\epsilon(n_{b}, n_N, Z)}{\partial n_N}=\mu_N^\mathrm{CLD}+\mu_N^\mathrm{shell}. \label{mu_N_shell}
\end{eqnarray}
Here, as in Section \ref{Sec_EqsSmooth}, the baryon chemical potential $\mu_b$ corresponds to the energy change
due to addition of a
baryon at fixed proton number density, $Zn_N$. 
Since $Z n_N$ is fixed,  
the added baryon is a neutron and $\mu_b$ can (again) be identified with the neutron chemical potential, $\mu_b=\mu_n$.
In Eqs.\ (\ref{mu_n_shell}) and (\ref{mu_N_shell}) 
$\mu_n^\mathrm{CLD}$ and $\mu_N^\mathrm{CLD}$ are corresponding chemical potentials
obtained from the CLD model
\begin{eqnarray}
    \mu_n^\mathrm{CLD}&=&
        \frac{\partial \epsilon^{\rm CLD}(n_b,n_{N},Z)}{\partial n_b},\\ 
    \mu_N^\mathrm{CLD}&=&\frac{\partial \epsilon^{\rm CLD}(n_b,n_{N},Z)}{\partial n_N},
\end{eqnarray}
while corrections, associated with the shell effects are
\begin{eqnarray}
    \mu_n^\mathrm{shell}&=&
        n_{N}\frac{\partial E^{\rm shell}(n_b,n_{N},Z)}{\partial n_b},
    \label{mu_n_shellCor}
\\
    \mu_N^\mathrm{shell}&=&E^{\rm shell}
        +n_{N}\frac{\partial E^{\rm shell}(n_b,n_{N},Z)}{\partial n_N}.
    \label{mu_N_shellCor}
\end{eqnarray}

Technically, to derive  mEOS in the form $\epsilon(P,\mu_n, Z)$, 
utilized
in Section \ref{Sec_EqsShell}, we employ numerical solvers to  obtain the parameters $n_{b}$, $n_N$ for given  $P$, $\mu_n$ and $Z$, in accordance with Eqs.\ (\ref{P_shell}) and (\ref{mu_n_shell}).

\subsection{Models of the shell effects and numerical implementation}\label{Sec_Shell_Models}

In this section, we 
discuss
the proton shell models 
used
in this study.
We recall that calculating  shell effects in the inner crust is a model-dependent problem. 
Given the uncertainties outlined below, we consider the results obtained with shell models employed in this paper as a reasonable first step and a proof-of-principle calculation, providing a foundation for subsequent calculations with more refined models.

All our shell models are primarily based on the supplementary tables presented in Ref.\
\cite{Pearson_ea18_bsk22-26,Pearson_ea19_ShellCorr_Errata}.%
\footnote{The tables were downloaded from  \href{https://doi.org/10.1093/mnras/stz800}{https://doi.org/10.1093/mnras/stz800 } on December  04, 2019.}
However, adapting these tables to our problem is not straightforward, and thus, we need to delve into certain technical details and assumptions to clarify their application.

Firstly, it is important to note that the tables presented as supplementary data in Ref.\ 
\cite{Pearson_ea19_ShellCorr_Errata} may not be directly applicable to our case. The reported 
values were obtained by minimizing the ETF energy density over $n_N$ at fixed $Z$ and $n_{b}$, 
which is not appropriate for accreted crust
(see Section \ref{Sec_Micr_Smooth}).
Nevertheless, given that the resulting EOS closely resembles the catalyzed case,  it seems 
reasonable  that actual shell energies can be well approximated by assuming that $E^{\rm shell}$ 
does not explicitly depend on $n_{b}$, but primarily depends on the cell size, i.e., $n_N$. 
Consequently, the shell corrections are expressed in the form 
\begin{equation}
    \Delta \epsilon^{\rm shell}=n_{N} E^{\rm shell}(n_{N},Z), 
  \label{shell2}
\end{equation}
where $E^{\rm shell}(n_{N},Z)$ can be obtained from the tables of Ref.\ \cite{Pearson_ea19_ShellCorr_Errata}.
Note that, in this approximation, shell corrections to the neutron chemical potential vanish, $\mu_n^\mathrm{shell}=0$ 
[see Equation (\ref{mu_n_shellCor})].

Secondly, it is worth mentioning that for nucleon interaction potential BSK24, applied in this work,
the tables in Ref.\ \cite{Pearson_ea19_ShellCorr_Errata} only provide shell energies for even $Z$.
However, to consider beta-capture and beta-emission reactions, we need information about energies of odd-$Z$ nuclei.
To analyze the role of the pairing effects, we consider two models for odd-$Z$ nuclei.
In the first model, referred to as ``Shell'',
the shell energies for
odd $Z$ are assumed to vanish.
In the second model, referred to as ``Shell+Pairing'', we set the shell energies for odd-$Z$ nuclei to be equal to the  pairing term.
The latter is estimated within the qualitative model suggested in \cite{Ludwig_ea73} and also used 
by Mackie and Baym \cite{MB77}:
\begin{equation}
E^{\rm shell}(n_{N},Z)=\frac{11 \mathrm{MeV}}{\sqrt{A(n_N)}}\mathrm{~ for \,\, odd \,\, }Z.
\end{equation}
Here $A(n_N)$ represents the dependence of the mass number $A$ on 
the number density of nuclei $n_N$, which, for simplicity, was adopted from the results of $Z$-fixed calculations (see below). 

Thirdly, we assume  that  shell corrections become negligible  above the proton drip density.
This implies that we set $E^{\rm shell}(n_{N},Z)=0$, if $n_{b}$ in the respective line of the table \cite{Pearson_ea19_ShellCorr_Errata} 
exceeds the proton drip density $n_{b}^{p,\mathrm{drip}}$. For simplicity, we assume the proton 
drip density to be the same for all $Z$, $n_{b}^{p,\mathrm{drip}}=0.073$~fm$^{-3}$ 
\cite{Pearson_ea18_bsk22-26}.
For some $Z$, we apply a smooth suppression of the function $E^{\rm shell}(n_{N},Z)$ as we approach 
the proton drip to avoid unrealistically sharp jumps, which could impact the pressure in our 
thermodynamically consistent approach [see Eqs.\ (\ref{P_shell})--(\ref{mu_N_shell})]. 
For numerical applications, we also eliminate some outliers from the tables in Ref.\ \cite{Pearson_ea19_ShellCorr_Errata}  and fit the remaining data on the shell energies 
as functions of $n_N$ separately for each $Z$.

Finally,  it is worth noting that the shell energy tables in \cite{Pearson_ea19_ShellCorr_Errata} also do not contain data for $Z<18$. For such nuclei, the shell energy was assumed to be zero, which 
most certainly
has a negligible effect on  our calculations
because formation of nuclei with $Z<18$ is blocked by high potential $\psi$ for  $Z=18$ nuclei (see Figure \ref{Fig_PsiProf_shell}).

We also employ the third (simplified) model as a sensitivity test for our results.
In this model we assume that the charge number $Z$ remains fixed, $Z=20$,
up to the proton drip, effectively mimicking very strong shell and pairing effects.
While, strictly speaking, the dependence of the shell energy on $n_N$
influences the results by affecting the pressure, we simplify this model by assuming $E^{\rm shell}(n_{N},Z=20)=0$. 
The results for the $Z$-fixed model are qualitatively aligned with both the Shell and Shell+Pairing 
models. The quantitative differences can be used to estimate uncertainties associated with shell 
effects (see Sections \ref{Sec_Res_Shell} and \ref{Sec_Heurist} for details).

%%%%%%%%%%%%%%%%%%%%%%%%%%%%%%%%%%%%%%%%%%%%%%%%%%%%%%%%%%%%%%%%%%%%%%%%%%%%%%%%%%%%%%%%%%%%%
\section{nHD models of FAC} \label{Sec_Res}
%%%%%%%%%%%%%%%%%%%%%%%%%%%%%%%%%%%%%%%%%%%%%%%%%%%%%%%%%%%%%%%%%%%%%%%%%%%%%%%%%%%%%%%%%%%%%

%%%%%%%%%%%%%%%%%%%%%%%%%%%%%%%%%%%%%%%%%%%%%%%%%%%%%%%%%%%%%%%%%%%%%%%%%%%%%%%%%%%%%%%%%%%%%
\subsection{Smooth CLD model} \label{Sec_Res_Smooth}
%\subsection{Smoothed CLD model} \label{Sec_Res_Smooth}
%%%%%%%%%%%%%%%%%%%%%%%%%%%%%%%%%%%%%%%%%%%%%%%%%%%%%%%%%%%%%%%%%%%%%%%%%%%%%%%%%%%%%%%%%%%%%
%
\begin{figure}[t!]
	\includegraphics[width=0.992\columnwidth]{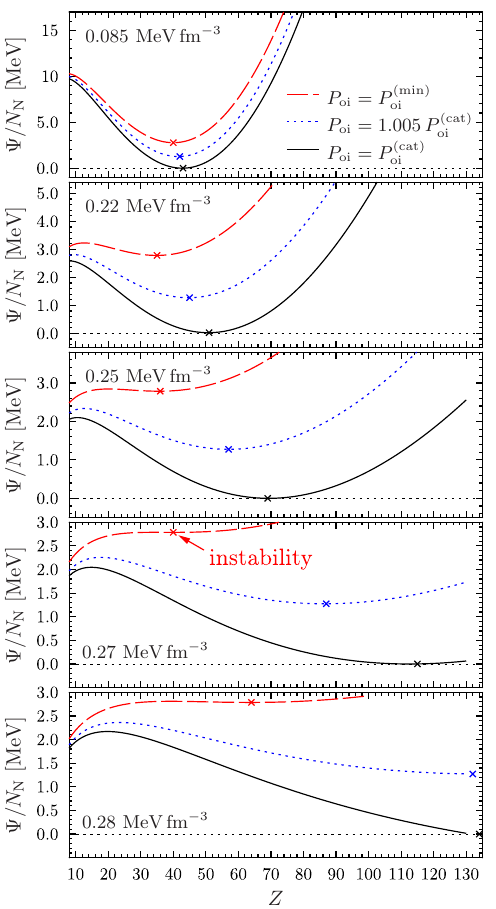}
	\caption{Profiles $\psi(Z)=\Psi/N_N$ for several values of pressure $P$ for the three  members of nHD EOS family, calculated
    with smooth CLD model (see text for details). 
    Crosses indicate minima
    or an inflection point (marked as instability) of the function $\Psi/N_N$. }
	\label{Fig_PsiProf_smooth}
\end{figure}

Let us begin the discussion of the results 
for the smooth CLD model by examining the profiles of the potential $\psi$ as a function of $Z$ for five values of pressure. The panels 
(from top to bottom)
correspond to $P=0.085$, $0.22$, $0.25$, $0.27$, and $0.28$~MeV\,fm$^{-3}$, respectively. In each panel, three members of the nHD EOS family are shown:
$P_\mathrm{oi}=P_\mathrm{nd}^\mathrm{(cat)}$ (solid line),
$P_\mathrm{oi}=1.005\,P_\mathrm{nd}^\mathrm{(cat)}$ (dotted line), and
$P_\mathrm{oi}=P_\mathrm{oi}^\mathrm{(min)}\approx 1.011\,P_\mathrm{nd}^\mathrm{(cat)}$ (dashed line).
Here $P_\mathrm{nd}^\mathrm{(cat)}$ is the pressure at the oi interface for catalyzed crust, which coincides with the neutron drip pressure (see \cite{Chamel_etal15_Drip} for discussion). Following \cite{GC21_HeatReleaze}, we indicate this by the subscript ``nd''.

In the smooth CLD model, the  nHD EOS for the entire crust can be  specified 
by setting the pressure $P_\mathrm{oi}$ at the outer-inner crust interface. Consequently, the solid line, corresponding to $P_\mathrm{oi}=P_\mathrm{nd}^\mathrm{(cat)}$, represents the $\psi$ profiles
for
catalyzed crust. 
At each pressure
the profiles exhibit profound minima, as clearly visible in
Figure \ref{Fig_PsiProf_smooth}. These minima correspond to the catalyzed crust composition 
($\psi=\mu_N=0$ at the minima; see Section \ref{Sec_EqsSmooth}),
indicating the absence of disintegration instability.
It is important to note, however, that all panels, except the 
top one, correspond to pressures where the pasta phases are more energetically favorable in the CLD model, as reported in Ref.\ \cite{DinhThi_ea21_CLD_pasta}. Hence, nuclei with large $Z$ ($Z>50$) are likely absent in the catalyzed crust, being shown here only for illustrative purposes.

The dotted line represents a slightly higher value of $P_\mathrm{oi}=1.005\,P_\mathrm{nd}^\mathrm{(cat)}$, which, however, is not 
sufficient to induce instability. The $\psi$ profiles still exhibit minima at all pressures, although these minima are less pronounced than for solid curves ($P_\mathrm{oi}=P_\mathrm{nd}^\mathrm{(cat)}$). Nonetheless, they still correspond to a stable composition. Therefore, the FAC model would require a larger $P_\mathrm{oi}$ to reach instability.

The dashed line corresponds to $P_\mathrm{oi}=P_\mathrm{oi}^\mathrm{(min)}\approx 1.011\,P_\mathrm{nd}^\mathrm{(cat)}$, the lowest $P_\mathrm{oi}$ value that leads to disintegration instability in the nHD inner crust. 
For this $P_\mathrm{oi}$, the  minima in the $\psi$ profiles become progressively shallower with increasing pressure and eventually disappear, transforming into an inflection point in the panel corresponding to $P=0.27$~MeV\,fm$^{-3}$ (the second panel from the bottom).
At this pressure, the nuclei disintegrate, allowing for 
the existence of stationary accreted crust.
However, it is worth noting that the neutron chemical potential at this point does not 
match the core $\mu_n$ (at the same pressure), 
and thus, the crust must be extended to higher pressures.
These layers can be called ``relic'' \cite{GC21_HeatReleaze}, 
since  the accreted nuclei do not penetrate into 
these layers  (they disintegrate earlier).
The relic layers can be
filled with spherical nuclei formed during the initial stages of accretion. 
Indeed, at a higher pressure, $P=0.28$~MeV\,fm$^{-3}$ (the bottom panel), 
a minimum in the $\psi(Z)$ profile for the dotted curve reappears, suggesting that stable relic crust can exist. Alternatively, the relic layers could be filled with pasta, or a layer of relic spherical nuclei could be followed by a pasta layer.%
%
%%%%%%
\footnote{To avoid any confusion, by ``spherical nuclei'' we mean, following, e.g., Ref.\ 
\cite{Pearson_ea18_bsk22-26} nuclei whose energy is calculated within the ETFSI approach using a 
spherically symmetric single-particle effective potential.}
%%%%%%
%
A detailed study of the structure of relic layers is beyond the scope of the present work.

For a smooth CLD model, the nHD solutions with $P_\mathrm{oi}>P_\mathrm{oi}^\mathrm{(min)}$ contain 
an unstable region where $\Psi/N_\mathrm{N}$ is a monotonically increasing function of $Z$. This 
region is located between the accreted and relic parts of the crust, indicating that the crustal 
model with $P_\mathrm{oi}>P_\mathrm{oi}^\mathrm{(min)}$ is thermodynamically inconsistent and 
cannot be applied to describe neutron stars \cite{GC20_DiffEq,GC21_HeatReleaze}. Therefore, the nHD 
model with $P_\mathrm{oi}=P_\mathrm{oi}^\mathrm{(min)}$ is the only possible stationary crust 
model, 
and it is referred 
to as simply the ``FAC model'' in what follows.

\begin{figure}[t!]
	\includegraphics[width=0.992\columnwidth]{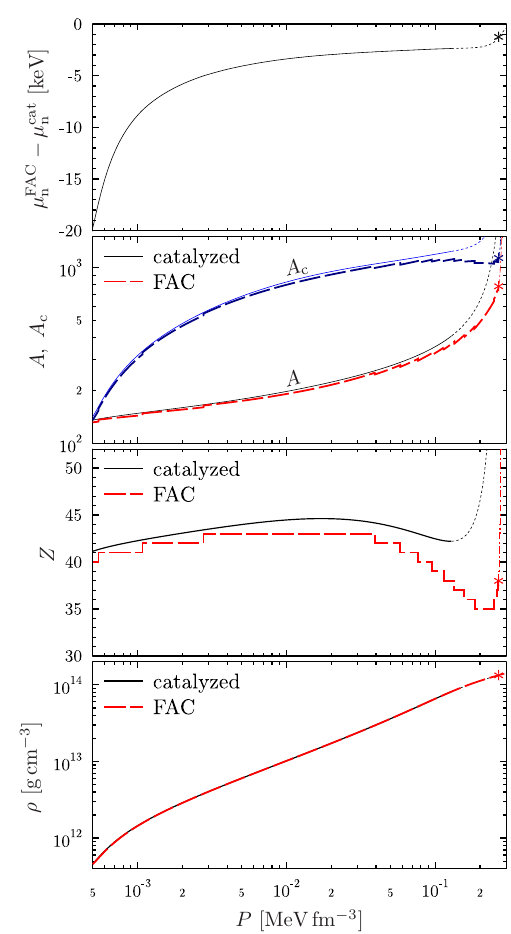}
	\caption{Crustal EOS and composition 
 for catalyzed crust and FAC in the smooth CLD model.
 In the figure $\rho=\varepsilon/c^2$ is the density and $A_c=n_b/n_N$ is the total number of nucleons per one nucleus.}
	\label{Fig_EOS_smooth}
\end{figure}

Figure \ref{Fig_EOS_smooth} compares two EOSs: the catalyzed crust and
the FAC model.
The density and composition of catalyzed crust  are shown by solid lines for $P<P_\mathrm{pasta}\approx 0.13$~MeV\,fm $^{-3}$ and  dotted lines for $P\ge P_\mathrm{pasta}$.
The dotted lines are
used to indicate that the region $P\ge P_\mathrm{pasta}$ may actually contain strongly nonspherical (pasta-like) structures \cite{DinhThi_ea21_CLD_pasta},  although  
the curves were 
calculated
assuming spherical nuclei.

The density and composition of the FAC model
are shown in Figure \ref{Fig_EOS_smooth} by thick dashes.
The instability takes place at 
$P_\mathrm{inst}\approx 0.27$~MeV\,fm$^{-3}$, marked with asterisks in each panel of Figure  \ref{Fig_EOS_smooth}.
As discussed above, $P_\mathrm{inst}$  is lower than the crust-core boundary and the crust should continue with  relic layers.
In Figure \ref{Fig_EOS_smooth}, the density and composition of relic layers are represented by thin dash-dotted lines, assuming that these layers consist of spherical nuclei.
It is worth noting that for the smooth CLD model, the composition of these layers can be unambiguously determined for a given $P_\mathrm{oi}$,
as the nHD equilibrium of relic layers allows us to apply formulas from Section \ref{Sec_EqsSmooth}. However, if shell effects are included, determining the composition of relic layers generally requires consideration of their formation history from the beginning of accretion process.

As seen from Figure \ref{Fig_EOS_smooth},
both density and composition profiles
of the FAC state closely resemble those of the catalyzed crust.
The fact that the instability occurs at a higher density than the crust-pasta transition, as predicted by \cite{DinhThi_ea21_CLD_pasta}, suggests that the thickness of the pasta layers may be affected by accretion. However, we defer a more detailed discussion of these effects and the role of the pasta in the accreted crust to subsequent studies.

The upper panel in Figure \ref{Fig_EOS_smooth} illustrates the difference in $\mu_{n}(P)$ between 
the catalyzed crust and FAC model. It is evident that the difference is quite small ($\lesssim 
20$~keV) and diminishes with increasing pressure. 
As in the other panels, the dotted 
part of the curve  is calculated assuming 
that the catalyzed crust is composed of spherical nuclei 
at such densities.

\subsection{CLD model with shell effects} \label{Sec_Res_Shell}
\begin{figure}
	\includegraphics[width=0.992\columnwidth]{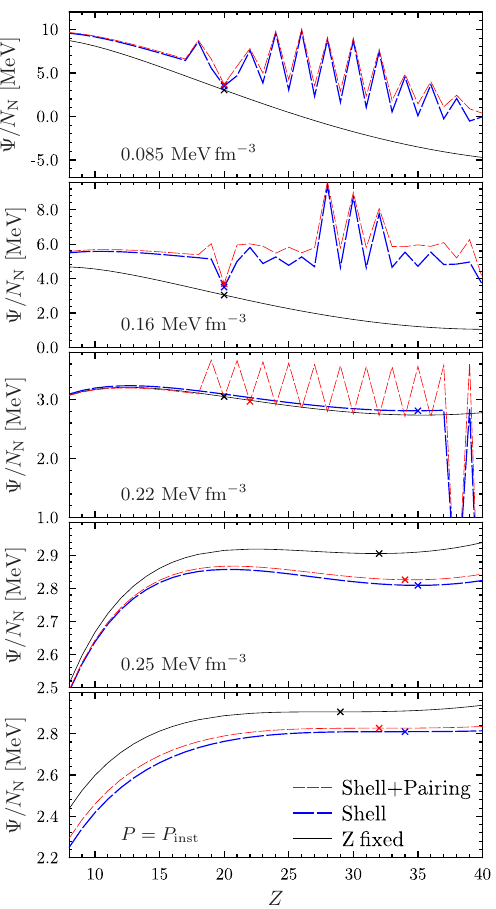}
	\caption{Profiles of the potential $\psi=\Psi/N_N$ for the nHD inner crust model, corresponding to $P_\mathrm{oi}=P_\mathrm{oi}^\mathrm{(min)}$, plotted for several values of pressure $P$ and three considered shell models.}
	\label{Fig_PsiProf_shell}
\end{figure}

As for the smooth CLD model, let us begin the discussion of the results by examining the profiles of the potential $\psi$, which are shown in Figure \ref{Fig_PsiProf_shell}.
In contrast to Figure \ref{Fig_PsiProf_smooth}, the lines in each panel correspond to different 
shell models (see Section \ref{Sec_Shell_Models}): $Z$-fixed (solid line), Shell (long dashes), and Shell+Pairing (short dashes). We have chosen $P_\mathrm{oi}$ to be equal to $P_\mathrm{oi}^\mathrm{(min)}$, which is specific for each model 
($P^\mathrm{(min)}_\mathrm{oi}\approx 0.432$~keV\,fm$^{-3}$ for Shell and Shell+Pairing models; 
$P^\mathrm{(min)}_\mathrm{oi}\approx 0.424$~keV\,fm$^{-3}$ for the $Z$-fixed model).
The panels, from top to bottom, correspond to $P=0.085$, 0.16, 0.22, 0.25 MeV\,fm$^{-3}$ and $P=P^\mathrm{inst}$.
The charge of the nuclei is denoted by a cross (for each model in each panel).
As expected, for Shell and Shell+Pairing models, it corresponds to a local  minimum of the potential $\psi$ in each panel. The  exception is the bottom panel, where the minimum disappears and becomes an unstable inflection point.
For the $Z$-fixed model, the charge is artificially fixed at $Z=20$ up to the proton drip (three 
upper panels).
That is, $Z=20$ does not correspond to the minimum of the $\psi$ potential.

In two upper panels ($P=0.085$ and $0.16$ MeV\,fm$^{-3}$)  a profound minimum of the $\psi$ potential is formed by the shell effects for both Shell and Shell+Pairing models. This guarantees the conservation of the nuclear charge number  since beta-capture and beta-emission reactions are not energetically favourable.
For the $Z$-fixed model $Z$ is fixed, thus for all considered models, $Z$ is equal to 20 at these 
pressures.

In the third panel ($P=0.22$ MeV\,fm$^{-3}$)  the charge number starts to differ.
For the $Z$-fixed model, it is kept equal to 20 by construction, while for the Shell model it 
evolves to $Z=35$. This is because the shell effects for low-$Z$ nuclei become small (according to 
the applied model), and the shell structure cannot prevent beta-reactions driven by the general 
trend of $\psi$ provided by the CLD part of the model. 
The proton drip has not yet been reached, and we retain pairing effects in the Shell+Pairing model.
They prevent a strong increase in
$Z$; however, $Z$ is increased to $Z=22$ due to a pair of beta emissions, which occur
between 
$P=0.16$ MeV\,fm$^{-3}$
and 
$P=0.22$ MeV\,fm$^{-3}$, when combination of the shell and pairing effects remove a 
$\psi$-potential barrier for the transition to $Z=21$ nuclei via beta emission. 

The fourth panel, $P=0.25$ MeV\,fm$^{-3}$, corresponds to the matter after the proton drip. Within our approximation, the shell and pairing  corrections vanish, and  $\psi(Z)$ becomes a smooth function of $Z$, as given by the CLD model.
The composition is driven to the minima located at $Z=32$, 34, and 35 for Shell, Shell+Pairing, and $Z$-fixed models, respectively.
The $\psi$ profiles become rather close for all the considered models.
Note, however, that they do not coincide exactly because the $\psi$ potential depends not only on $P$, but also on $\mu_n$, which is specific for each of the considered shell models due to differences 
in $P_\mathrm{oi}^\mathrm{(min)}$ and
in the respective EOSs. 
This feature leads to differences in nuclear evolution after the proton drip (see Figure  \ref{Fig_EOS_shell} below).

In the bottom panel, where $P=P^\mathrm{inst}$, there are no minima for all models; 
the
``optimal'' $Z$
correspond
to the inflection point. Nuclei become unstable and disintegrate through a sequence of beta 
captures, driving them 
to lower $Z$ and $\psi$.

\begin{figure}
	\includegraphics[width=0.992\columnwidth]{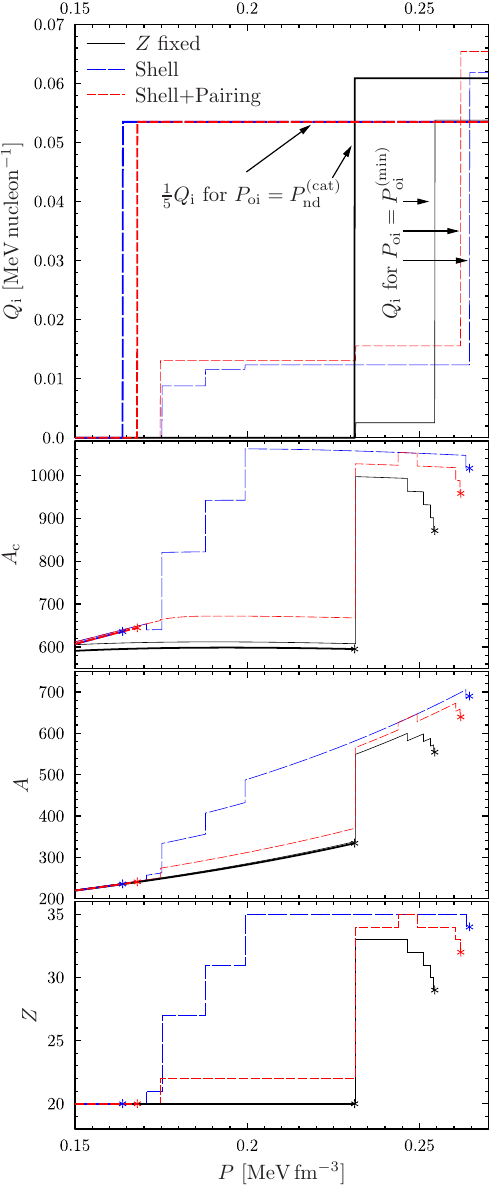}
	\caption{
 Accumulated heat per accreted baryon 
 $Q_i$ and composition ($Z$, $A$, and  the total number of nucleons $A_\mathrm{c}$ per one nucleus) versus $P$ for nHD crust 
 are shown for
 Shell, Shell+Pairing, and $Z$-fixed models.
 For each model, $Q_i$ and composition
 are shown for the two values of the pressure at the oi interface: $P_\mathrm{oi}=P_\mathrm{oi}^\mathrm{(min)}$ (thin lines) and $P_\mathrm{oi}=P_\mathrm{nd}^\mathrm{(cat)}$ (thick lines).}
	\label{Fig_EOS_shell}
\end{figure}

Let us now turn to discussing the evolution of nuclei
in the inner crust.
For $Z$-fixed and Shell+Pairing models, $Z$ at the oi interface starts from the value $Z=20$ for all the considered members of the nHD family of crust models, corresponding to different
$P_\mathrm{oi}$. 
These nuclei are formed by reactions in the outer crust and at the oi interface.
Subsequent compression up to 
$P\lesssim0.15$~MeV\,fm$^{-3}$ 
does not lead to beta-reactions and the charge number 
remains
to be $Z=20$. No heat is released.
The 
evolution consists 
of 
continuous growth of 
nuclear mass numbers
$A$ and $A_c$ with 
pressure, $P$.
Neutrons required for this growth are provided by diffusion/superfluid flow from 
the higher-density regions.
The evolution for the Shell model is generally the same, but
$Z=21$ nuclei are formed at the oi interface and converted into $Z=20$ nuclei by electron capture 
at a slightly higher pressure. This is a clear artifact of neglecting the pairing effects in that 
model. 
We expect that it does not significantly affect our results. In particular, the electron capture does not lead to  any energy release, because it takes place exactly at the threshold. 

Nuclear evolution in the inner layers of the inner crust is more interesting (see Figure \ref{Fig_EOS_shell}). 
Panels in that figure, from top to bottom, present the accumulated 
heat $Q_\mathrm{i}$ in the inner crust (per accreted baryon, not including the heat released at the oi interface),  
as well as $A_{c}$, $A$, and $Z$ as functions of $P$. Each panel contains two types of curves for each model (Z-fixed, Shell, and Shell+Pairing):
the thin curve shows the evolution
for $P_\mathrm{oi}=P^\mathrm{(min)}_\mathrm{oi}$, 
while the thick curve is for $P_\mathrm{oi}=P^\mathrm{(cat)}_\mathrm{nd}$.
In the lower panels, all curves
coincide at low pressure.
In the upper panel, the energy release for $P_\mathrm{oi}=P^\mathrm{(cat)}_\mathrm{nd}$ is divided 
by a factor of 5 to fit the scale of the plot. 
The instability points are shown by asterisks (except in the upper panel).

Let us first consider the case
$P_\mathrm{oi}=P^\mathrm{(cat)}_\mathrm{nd}$. Then, for Shell and Shell+Pairing models, the instability occurs for $Z=20$ nuclei
before the proton drip. 
This is 
due to
the high value of $P_\mathrm{oi}$, for which
the general trend in the $\psi(Z)$ dependence becomes so strong that the shell effects fail to 
prevent beta captures.
For the $Z$-fixed model, the instability occurs exactly at the proton drip point, simply because we do not allow $Z$ to vary at lower $P$ in this model.
All the heat $Q_\mathrm{i}$ is released at the instability point, resulting in a jump in the $Q_\mathrm{i}(P)$ dependence, as shown in the upper panel.

For $P_\mathrm{oi}=P^\mathrm{(min)}_\mathrm{oi}$, the nuclear evolution 
for the considered shell models is somewhat different.
However, the general trend is a growth of $Z$ up to $33-35$ with subsequent decrease before the onset of instability. The increase in $Z$ is accompanied by heat release; however, the released heat at the instability is dominant for all models.

Let us discuss the details of nuclear evolution, starting with the Shell model.
The first electron emission takes place at $P\approx 0.17$~MeV\,fm$^{-3}$, 
when the shell effects for nuclei with $Z=20$ become small and cannot 
prevent beta capture. 
The absence of pairing correction in this model allows for the formation of $Z=21$ nuclei. 
Subsequent electron captures are associated with a
further decrease  of the shell corrections, leading to gradual
disappearance of local minima of the function $\psi(Z)$. Finally, at $P\approx 0.20$~MeV\,fm$^{-3}$, the shell corrections vanish for low-$Z$ nuclei, and the nuclear charge arrives at a minimum of $\psi(Z)$, determined by the smooth CLD model. 
The negative shell energy corrections for high-$Z$ nuclei constitute the global minimum of $\psi(Z)$; however, this minimum remains 
unattainable.
This pattern stays the same with a subsequent increase in pressure, and the proton drip does not leave any imprints on the evolution, since the local shape of $\psi(Z)$ is already driven by the CLD model.
The decrease of $Z$ before the instability onset is associated with changes in the $\psi(Z)$ shape 
with growing $P$.

Within the Shell+Pairing model, the decrease 
of shell corrections at $P\approx 0.17$~MeV\,fm$^{-3}$ does not affect the evolution because the 
local minimum at $Z=20$ is well defined by the pairing correction. Subsequent compression leads to 
a pair of electron
emissions and the associated energy release from the second capture. 
However, the pairing correction adopted in the Shell+Pairing model is strong enough to form a local 
minimum at $Z=22$ and prevent subsequent electron emissions until the proton drip takes place. At 
the proton drip, we suppress the pairing corrections along with shell corrections, and the 
$\psi(Z)$ profile becomes determined by the CLD model. The nuclear charge reaches a minimum of 
$\psi(Z)$ potential, located at $Z=34$. During subsequent compression, $Z$ follows  the position of 
the minimum (see the second from the bottom panel in  Figure \ref{Fig_PsiProf_shell}).

The $Z$-fixed model is applied as a sensitivity test for shell effects. It prevents any evolution of the nuclear charge $Z$ until the proton drip point. For higher pressure, as in the Shell and Shell+Pairing models, the $\psi(Z)$ profile becomes determined by the CLD model, which drives the subsequent evolution. 

%%%%%%%%%%%%%%%%%%%%%%%%%%
\begin{figure}
	\includegraphics[width=0.992\columnwidth]{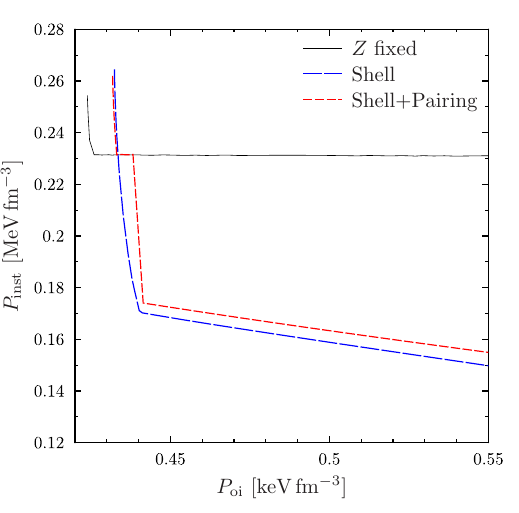}
	\caption{$P_\mathrm{inst}$ vs $P_\mathrm{oi}$ for Shell (short dashes), Shell+Pairing (long dashes) and $Z$-fixed (solid line) models. }
	\label{Fig_Pinst_vs_Poi}
\end{figure}
%%%%%%%%%%%%%%%%%%%%%%%%%%

Figure \ref{Fig_Pinst_vs_Poi} presents the dependence of $P_\mathrm{inst}$ on $P_\mathrm{oi}$ for the considered models. 
One can observe that $P_\mathrm{inst}$ decreases with  an increase in $P_\mathrm{oi}$  for all models.

Let us begin the discussion of this dependence with the $Z$-fixed model.
As seen in Figure \ref{Fig_EOS_shell}, for $P_\mathrm{oi}=P_\mathrm{oi}^\mathrm{(min)}$, the instability occurs after the proton drip. 
An increase in $P_\mathrm{oi}$ 
shifts
the instability point to lower pressure, closer to the proton drip, located at $P\approx 0.231$~MeV\,fm$^{-3}$. 
Finally, at $P_\mathrm{oi}\approx 0.425$~keV\,fm$^{-3}$, the instability starts at the proton drip.
In the $Z$-fixed model, we do not allow $Z$ to evolve before the proton drip. As a result, the instability cannot occur at a lower pressure. Consequently, for  $P_\mathrm{oi}> 0.425$~keV\,fm$^{-3}$, the instability always takes place at the proton drip point (some fluctuations in Figure \ref{Fig_Pinst_vs_Poi} are numerical noise).
This behavior is an artifact of the simplified $Z$-fixed approach, prompting us to switch to more realistic models.

For the Shell model, as we discussed above, even for $P_\mathrm{oi}=P_\mathrm{oi}^\mathrm{(min)}$, the nuclear charge starts to evolve before the proton drip point, and the proton drip does not affect the evolution. This statement holds true for higher  $P_\mathrm{oi}$ values, and $P_\mathrm{inst}(P_\mathrm{oi})$ is a monotonically decreasing function with no peculiarities at the proton drip point. The break at $P_\mathrm{oi}\approx 0.44$~keV\,fm$^{-3}$ is associated with a change in the nuclear evolution behavior: for lower $P_\mathrm{oi}$, the nuclear charge starts to grow (above some pressure), and the instability occurs for $Z>20$, while for higher $P_\mathrm{oi}$, the instability occurs for $Z=20$ 
nuclei without any charge evolution before it (see the  thick line for $P_\mathrm{oi}=P_\mathrm{nd}^\mathrm{(cat)}$ and the thin line for $P_\mathrm{oi}=P_\mathrm{oi}^\mathrm{(min)}$ in Figure \ref{Fig_EOS_shell}).

The behavior of $P_\mathrm{inst}(P_\mathrm{oi})$ is a bit more complicated for the Shell+Pairing model.
As with the Shell model, for $P_\mathrm{oi}=P_\mathrm{oi}^\mathrm{(min)}$, the nuclear charge starts to evolve before the proton drip, but this evolution is not too strong; pairing effects prevent $Z$ from growing above $Z=22$ (see Figure \ref{Fig_EOS_shell}). At the proton drip point, we remove shell and pairing corrections, and the nuclear charge increases.

Increasing $P_\mathrm{oi}$ results in a decrease in $P_\mathrm{inst}$, while qualitatively nuclear evolution remains unchanged:
The nuclear charge increases at the proton drip point and (generally) evolves a little further before the instability onset. 
However, at $P_\mathrm{oi}\approx 0.433$~keV\,fm$^{-3}$, the instability point reaches the proton drip, where instability leads to the disintegration of $Z=22$ nuclei. Subsequent growth of $P_\mathrm{oi}$ up to $0.44$~keV\,fm$^{-3}$ does not affect $P_\mathrm{inst}$, because the pairing effects prevent disintegration of $Z=22$ nuclei before the proton drip. 
However, as with the Shell model, for $P_\mathrm{oi}>0.44$~keV\,fm$^{-3}$, the instability occurs 
before the proton drip in the form of disintegration of $Z=20$ nuclei, without an increase of nuclear charge before the instability.
In this high-$P_\mathrm{oi}$ region, the value of $P_\mathrm{inst}$ is slightly higher for the Shell+Pairing model than for the Shell model because the pairing correction provides additional stabilization for $Z=20$ nuclei.

%%%%%%%%%%%%%%%%%%%%%%%%%%%%%%%%%%%%%%%%%%%%%
\section{Constraints on $P_\mathrm{oi}$}
\label{Sec_Poi}
%%%%%%%%%%%%%%%%%%%%%%%%%%%%%%%%%%%%%%%%%%%%%

In the previous section, we considered nHD models of the accreted crust, parametrized by the pressure at the oi interface, $P_\mathrm{oi}$. In this section, 
we discuss constraints for this quantity in the FAC state.

Firstly, as discussed in Ref.\ \cite{GC20_DiffEq}, there is a mechanism for nuclei disintegration, which is required in the stationary FAC state to prevent the accumulation of nuclei supplied through accretion.
Within the nHD approach, this mechanism naturally arises at 
$P_\mathrm{oi}>P_\mathrm{oi}^\mathrm{(min)}$ in the form of instability, associated with the disappearance of the local minimum of the potential $\psi(Z)$ (see Figures \ref{Fig_PsiProf_smooth} and \ref{Fig_PsiProf_shell}). This establishes a lower bound for $P_\mathrm{oi}$.

But what about the upper bound?

For the smooth CLD model, it arises naturally when one assumes that nuclei are spherical throughout the entire crust. In this case, as discussed in Section \ref{Sec_Res_Smooth},  the relic crust can be unambiguously constructed using formulas from Section \ref{Sec_EqsShell}. The parameter $C=\mu_N/\mu_n$, see Eq.\ (\ref{cat3}), should remain constant  throughout the entire inner crust, including the relic region.
However, the relic region is stable 
and can connect the accreted crust and core
only when  $P_\mathrm{oi}=P_\mathrm{oi}^\mathrm{(min)}$ (see Figure \ref{Fig_PsiProf_smooth} and the corresponding discussion). 
For $P_\mathrm{oi}>P_\mathrm{oi}^\mathrm{(min)}$, there exists a pressure region, where the beta-equilibrium  
equation (\ref{cldm}) cannot be satisfied because $\psi(Z)$ 
increases monotonically with an increase in $Z$. 
Hence, a crust cannot exist for $P_\mathrm{oi}>P_\mathrm{oi}^\mathrm{(min)}$ and $P_\mathrm{oi}=P_\mathrm{oi}^\mathrm{(min)}$ represents the only possible value for $P_\mathrm{oi}$ in the nHD FAC for the smooth CLD model.
Note, however, 
that this conclusion should be reconsidered if we allow for nonspherical nuclei in relic crustal layers. 
We do not explore this possibility here,
but we anticipate that it may allow for slightly higher values of $P_\mathrm{oi}$, although an upper limit for $P_\mathrm{oi}$ should still exist.

As demonstrated in the previous section, for  more realistic nHD models that include shell effects, $P_\mathrm{inst}$ can be calculated for a given  $P_\mathrm{oi}$, and  the general trend is that $P_\mathrm{inst}$ decreases with $P_\mathrm{oi}$ (see Figure \ref{Fig_Pinst_vs_Poi}).
As discussed in the Introduction, $P_\mathrm{oi}$ and the composition of the relic part located at 
$P>P_\mathrm{inst}$, can, in principle, 
be calculated by considering crustal evolution from the very beginning of the accretion process. This is a very complicated (and model-dependent) problem that is beyond the scope of this work.
Instead, here we constrain the allowed $P_\mathrm{oi}$ region based on the requirement that 
the relic crust, which connects the instability point with the stellar core in a thermodynamically consistent way,%
%
%%%%%%%%
\footnote{
Thermodynamic consistency requires that:
1) the relic part of the crust must be mechanically stable at each point;
2) the nHD condition must be satisfied;
3) the pressure and neutron chemical potential must be continuous at both the instability point and  crust-core interface.
}
%%%%%%%%%
%
should be allowed at least for some compositions.
Indeed, if the region {\it before} the instability point cannot be connected with the core, the 
respective crustal model cannot occur in a real neutron star and should be disregarded.

If one attempts to constrain $P_\mathrm{oi}$ by formally applying the shell models from this work 
(see Section \ref{Sec_Micr_Shell}),  one should arrive at the conclusion that $P_\mathrm{oi}$ 
cannot exceed $P_\mathrm{oi}^\mathrm{(min)}$. 
Namely, within these models the shell corrections are absent after the proton drip, which is located lower than the crust-core boundary. 
Consequently, the part of the crust between the proton drip and the core is governed by the smooth CLD model.
As discussed above, for the smooth CLD model the nHD crust is determined by the constant $C$, and 
there exists one and only one value of this constant 
allowing one to  connect the instability point and the core with a stable relic crust, and this 
value corresponds to $P_\mathrm{oi}=P_\mathrm{oi}^\mathrm{(min)}$. 
However, this conclusion  relies heavily on the simplifications made in our shell models, therefore it might not be very reliable.

To establish a more reliable upper bound for 
$P_\mathrm{oi}$, we impose a less stringent requirement: the relic part, starting from the instability point, should remain stable up to the proton drip point for some composition of the relic part.

It is rather difficult to check for this condition numerically, even if we restrict ourselves to the assumption that the relic crust region is composed of spherical nuclei.
In particular, the result evidently depends on the applied shell model.
However, as anticipated, the general trend is that the lower the value of $P_\mathrm{inst}$, the more challenging it becomes to identify a relic crust composition. 
The reason is straightforward: the relic region becomes thicker, and the slope of the smooth CLD 
part of the $\psi(Z)$ function becomes larger (more unstable) .

We have verified that a stable relic crust does not exist for the Shell model with $P_\mathrm{oi}=P_\mathrm{nd}^\mathrm{(cat)}$. Therefore, we propose using $P_\mathrm{nd}^\mathrm{(cat)}$ as a reference upper bound for $P_\mathrm{oi}$ for $^{56}$Fe ash.
We do not attempt to determine the upper bound for the Shell model more precisely because it appears to be model-dependent. 
We refrain from establishing an upper bound for the Shell+Pairing model, which relies on rather artificial accounting for pairing corrections that would clearly affect the numerical results.
Finally, for the $Z$-fixed model the instability cannot occur before the proton drip by construction. 

Finalizing this section, we should stress that the above discussion is based on the assumption that the relic part of the crust contains only spherical nuclei. If it is (at least partially) composed of nonspherical nuclear clusters (pasta phases), the constraints should be reconsidered, but we leave this problem beyond the scope of this work.

%%%%%%%%%%%%%%%%%%%%%%%%%%%%%%%%%%%%%%%%%%%%%%%%%%%%%%%
\section{Energy release in deep layers of the nHD crust 
as function of $P_{\rm inst}$}
%%%%%%%%%%%%%%%%%%%%%%%%%%%%%%%%%%%%%%%%%%%%%%%%%%%%%%%
\label{Sec_Qdeep}

 %%%%%%%%%%%%%%%%%%%%%%%%%%%%%%%%%
\begin{figure}
	\includegraphics[width=0.992\columnwidth]{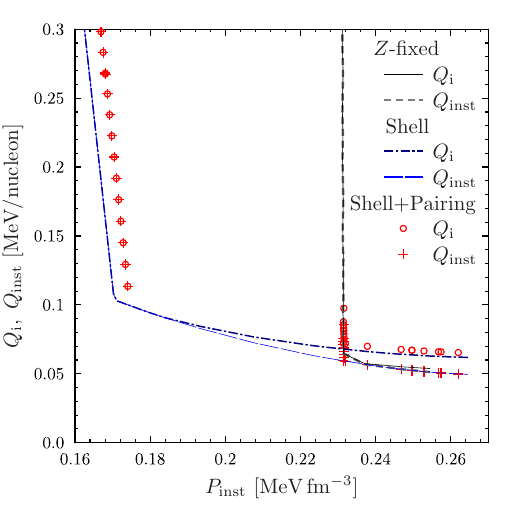}
	\caption{Heat release in the inner crust $Q_\mathrm{i}$ and at the instability point $Q_\mathrm{inst}$ as function of the instability pressure  $P_\mathrm{inst}$  for  $Z$-fixed, Shell, and Shell+Pairing models.
 }
	\label{Fig_Q_Pinst}
\end{figure}
%%%%%%%%%%%%%%%%%%%%%%%%%%%%%%%%%

As discussed in Section \ref{Sec_Res_Shell}, the pressure $P^\mathrm{inst}$ can be calculated as a function of $P_\mathrm{oi}$for a given shell model. 
Thus, $P^\mathrm{inst}$ can be used to parametrize the family of nHD models instead of 
$P_\mathrm{oi}$. 
Figure \ref{Fig_Q_Pinst} illustrates this point by demonstrating 
$Q_\mathrm{i}$  as a function of $P_{\rm inst}$.
Long and short dashes represent the Shell and $Z$-fixed models, while open circles denote the Shell+Pairing model.
Figure \ref{Fig_Q_Pinst} also displays the heat release at the instability, $Q_\mathrm{inst}$. 
Dash-dotted line, short dashes,  and crosses present  $Q_\mathrm{inst}$ for
the Shell, $Z$-fixed, and Shell+Pairing models, respectively.

The difference between  $Q_\mathrm{i}$ and  $Q_\mathrm{inst}$ is associated with the heat released in the inner crust before the instability, $P<P_{\rm inst}$. 
However, for all considered models, this heat release takes place in the very deep layers ($P>0.1$~MeV\,fm$^{-3}$), and for  $P_\mathrm{inst}<0.18$~MeV\,fm$^{-3}$, all the heat is released at the instability point for both Shell and Shell+Pairing models. 
The reason has been already discussed in Section \ref{Sec_Res_Shell}: at low enough $P_\mathrm{inst}$, the instability occurs for $Z=20$ nuclei without any energy release in the intermediate layers of the inner crust 
(i.e., between the oi interface and the instability point).

For the $Z$-fixed model, the lowest possible value of $P_\mathrm{inst}$ corresponds, by construction, to the proton drip. 
If the instability occurs at the proton drip, all heat is released at that point.

For high instability pressures, 
$P_\mathrm{inst}>0.24$~MeV\,fm$^{-3}$,
$Q_\mathrm{inst}$  becomes the same for all shell models applied in this work.
Indeed, for such $P_{\rm inst}$ the instability occurs after the proton drip, where all shell corrections for our models are set to zero (see Section \ref{Sec_Micr_Shell}).
Correspondingly, EOS near $P=P_{\rm inst}$ should be fully determined by the smooth CLD model.
In particular, immediately at the instability point one should satisfy the conditions (\ref{inst1}) and (\ref{inst2}) (see Section \ref{Sec_Algorithm} for details).
Imposing them, we can determine the parameters of the CLD model, namely $Z$ and $\mu_n$, and eventually calculate the potential $\psi$ at the instability point.
Because all incoming nuclei disintegrate at $P=P_{\rm inst}$, the energy release (per nucleus) there is equal to the potential $\psi(P_\mathrm{inst})$, being the same for all models
\cite{GKC_psi21}.
Since fusion reactions are absent for ashes composed of $^{56}$Fe,
the number of disintegrating nuclei should be the same as the number of 
nuclei supplied by the accretion.
That is, the energy release per accreted nucleon should be given by
$Q_\mathrm{inst}=\psi(P_\mathrm{inst})/A_\mathrm{ash}$ \cite{GKC_psi21}, being the function of $P_\mathrm{inst}$
($A_\mathrm{ash}=56$ is the number of nucleons per one nuclei in the ash).

\subsection{Minimal energy release}
\label{Sec_MinQ}

For each shell model, one can determine $Q_\mathrm{inst}^\mathrm{(min)}$ -- the minimum possible $Q_\mathrm{inst}$.
According to Figure \ref{Fig_Q_Pinst}, this corresponds to the highest possible pressure, $P^\mathrm{(max)}_\mathrm{inst}$, and numerically $Q_\mathrm{inst}^\mathrm{(min)}$ is the same for all considered shell models, $Q_\mathrm{inst}^\mathrm{(min)}\approx 0.05$~MeV per accreted nucleon. 

This latter statement should hold true for any shell model added on top of a given smooth CLD model, provided that shell effects are negligible at high pressures.
The proof of this crucial statement is, in fact, straightforward:
if shell effects are neglected at high pressures,  the nuclear physics there is  specified
by the smooth model% 
%
%%%%%%%
\footnote{It can be either a CLD model (as in the present work) 
or, for example, a more sophisticated smooth model based on the ETF calculations 
\cite{Pearson_ea18_bsk22-26,Pearson_ea19_ShellCorr_Errata}.
}.
%%%%%%%
As previously stated, in the smooth model the amount of energy release is determined by $P_\mathrm{inst}$. Thus, the function $Q_\mathrm{inst}(P_\mathrm{inst})$ and its minimum, $Q_\mathrm{inst}^\mathrm{(min)}$, are not affected by the shell 
effects.

In principle, $Q_\mathrm{inst}^\mathrm{(min)}$  can depend on the applied smooth model. The detailed analysis of this dependence is left for future work.
Here we would like to point out
that for another 
Skyrme-type nuclear potential (Sly4 \cite{Chabanat_ea98_SLY4}), the smooth CLD model leads to a very similar value for the minimal heat release ($0.04$~MeV per accreted nucleon, see supplementary material in Ref.\ \cite{GC21_HeatReleaze}). This suggests that $Q_\mathrm{inst}^\mathrm{(min)}\sim (0.04\div 0.05)$~MeV per accreted nucleon  can be considered a rather model-independent estimate.

%%%%%%%%%%%%%%%%%%%%%%%%%%%%%%
\section{Heuristic predictions for the nHD inner crust}
\label{Sec_Heurist}
%%%%%%%%%%%%%%%%%%%%%%%%%%%%%%

In this work, we analyze the nHD accreted crust for the three CLD+sh models described in Section \ref{Sec_Micr_Shell}.
We believe that these models provide a reasonable framework for the up-to-date description of the nuclear physics in the deepest layers of the inner crust.
However, for the shallow regions of the inner crust, where the number density of unbound neutrons is negligible, 
a better approach exists. 
It is based on the (theoretical) atomic mass tables (AMTs) (see Refs.\ \cite{lau_ea18,SC19_MNRAS}, for applications of AMTs to accreted crust modeling).
Currently available AMTs are constructed using detailed HFB calculations (e.g., \cite{Goriely_ea10,Goriely_ea_Bsk22-26,Grams_ea23_3D_HFB}),  finite-range droplet macroscopic model (FRDM; \cite{FRDM95,FRDM12}), 
machine learning/statistical methods \cite{Shelley_Pastore21_massmodel,Stopani_ea23} etc.
The description of the shallow crust, obtained with 
AMTs is
expected to be more accurate and reliable than 
that achieved 
with CLDM+sh models based on the 
ETFSI approach. 

In this section, we will assume that we have a hypothetical {\it advanced shell model} that, on the one hand, reproduces the AMT-based mEOS for the shallow region of the inner crust and, on the other hand, reduces to one of our three CLD+sh models for the deepest inner crust layers. 
Note that we will not make any assumptions about the form of
shell corrections in the crustal region with intermediate densities, 
except for assuming that their behavior is physically reasonable (smooth and continuous, see below for more precise applicability conditions).
Our goal will be to address the question of how a more realistic description of the shallow regions can affect the main parameters of the theory, such as the values of the pressures $P_{\rm oi}$ and $P_{\rm inst}$.

To answer this question, we will follow the approach based on the energy conservation law.
The approach requires almost no additional calculations, provided that the nHD inner crust has been 
preanalyzed making use of a CLD+sh model.
As a byproduct, we will demonstrate that the considered nHD properties do not depend on the details of the behavior of shell corrections in the intermediate layers of the inner crust.

The starting point for our energy-based 
approach is one of the central results of Ref.\ \cite{GC21_HeatReleaze}. 
In that reference it was shown that the
heat release $Q_\mathrm{i}$ can be calculated as a function of $P_\mathrm{oi}$ by using AMT and analyzing only nuclear reactions in the outer crust and at the oi interface.
The resulting function $Q^\mathrm{AMT}_\mathrm{i}(P_\mathrm{oi})$
is independent of the inner crust physics and hence
should be valid for the advanced shell model. Below $Q^\mathrm{AMT}_\mathrm{i}(P_\mathrm{oi})$ is 
assumed to be known (precalculated; see Refs.\ \cite{GC21_HeatReleaze,SGC_OC21} for examples of 
such a calculation).

%%%%%%%%%%%%%%%%%%%%%%%%%%%%%%%%%%%%%%%%%%%%%%%%%%%    
\subsection{Heuristic constraints on $P_\mathrm{oi}$}
\label{Sec_HeuristPoi}
%%%%%%%%%%%%%%%%%%%%%%%%%%%%%%%%%%%%%%%%%%%%%%%%%%
In Ref.\ \cite{GC21_HeatReleaze}, we 
constrained $P_\mathrm{oi}$ by imposing a condition that $Q^\mathrm{AMT}_\mathrm{i}(P_\mathrm{oi})>0$.
The corresponding lower bound for the pressure was denoted as $P_\mathrm{oi}^{(0)}$.
Here we suggest applying a tighter constraint: $Q^\mathrm{AMT}_\mathrm{i}(P_\mathrm{oi})>Q_\mathrm{inst}^\mathrm{(min)}$, where $Q_\mathrm{inst}^\mathrm{(min)}$ represents the minimal energy release at the instability.
As discussed in Section \ref{Sec_MinQ}, for BSK24 and SLy4 models $Q_\mathrm{inst}^\mathrm{(min)}$ can be estimated as $Q_\mathrm{inst}^\mathrm{(min)}\sim (0.04\div 0.05)$~MeV per accreted nucleon, provided that shell corrections are negligible in the deepest layers of the inner crust.
By using the dependence 
$Q^\mathrm{AMT}_\mathrm{i}(P_\mathrm{oi})$
calculated in Ref.\ \cite{GC21_HeatReleaze} (see inset in Figure \ref{Fig_Pinst_vs_Poi_cor}, $Q_\mathrm{inst}^\mathrm{(min)}\sim (0.04\div 0.05)$~MeV is filled in gray in the inset) for the BSK24 HFB mass tables \cite{Goriely_ea_Bsk22-26}, we obtain $P_\mathrm{oi}\ge
0.46$~keV\,fm$^{-3}$.
However, this constraint is only slightly stronger (by $\sim 2\%$)  than the originally suggested lower bound $P_\mathrm{oi}^{(0)}$ 
\cite{GC21_HeatReleaze}.
Note, however, that a detailed analysis within the advanced shell model will likely lead to the same result.

As discussed in Section \ref{Sec_MinQ}, $Q_\mathrm{inst}^\mathrm{(min)}$ is determined by the CLD model and thus should not depend on the initial composition.
Thus, the approach discussed in this subsection is likely applicable to accreted crust with a 
multicomponent composition of initial ashes.
Note, however, that in the latter case, 
noticeable fraction of the heat can be released in the intermediate layers
of the inner crust \cite{SGC22,SGC23_compos}, so that the constraint should be applied to the residual part of the heat release, which also can be estimated within the AMT approach \cite{SGC22,SGC23_compos}.

%%%%%%%%%%%%%%%%%%%%%%%%%%%%%%%%
\subsection{Heuristic predictions for $P_\mathrm{inst}(P_\mathrm{oi})$}
\label{Sec_HeuristPoi_vs_Pinst}
%%%%%%%%%%%%%%%%%%%%%%%%%%%%%%%%

%%%%%%%%%%%%%%%%%%%%%%%%%%%%%%%%
\begin{figure}
	\includegraphics[width=0.992\columnwidth]{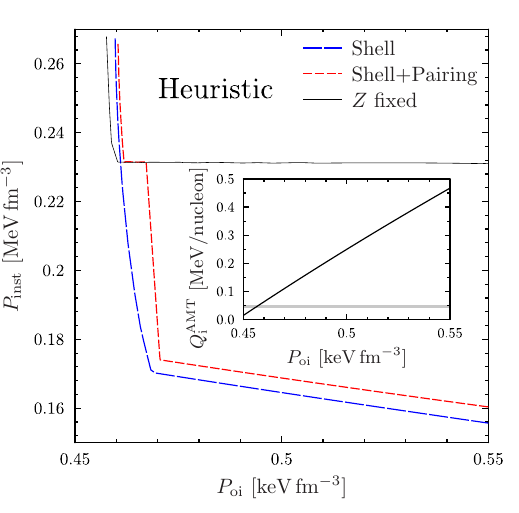}
	\caption{Heuristic prediction for the dependence 
 $P_\mathrm{inst}(P_\mathrm{oi})$
 for the advanced Shell (short dashes), Shell+Pairing (long dashes) and $Z$-fixed (solid line) models (see text for details). 
	The inset indicates $Q_\mathrm{i}^{\rm AMT}(P_\mathrm{oi})$ dependence as it was calculated in Ref.\ \cite{GC21_HeatReleaze} for the BSK24 HFB mass tables \cite{Goriely_ea_Bsk22-26}; $Q_\mathrm{inst}^\mathrm{(min)}\sim (0.04\div 0.05)$~MeV is filled in gray.}
 \label{Fig_Pinst_vs_Poi_cor}
\end{figure}
%%%%%%%%%%%%%%%%%%%%%%%%%%%%%%%%

Here we extend the approach from Section \ref{Sec_HeuristPoi} in order to 
heuristically predict the dependence of $P_\mathrm{inst}(P_\mathrm{oi})$ 
for the advanced shell model.
As before,  we employ the energy conservation law.
In addition, we make use of the assumption of constancy of $Z$ for the majority of the inner crust, up to the innermost regions, where the advanced shell model aligns with the respective CLD+sh model.
This assumption is supported by our calculations (see Sections \ref{Sec_Res_Shell} and \ref{Sec_Qdeep}) and can be considered as an applicability condition for the approach proposed in this subsection.

As pointed out above, $Q^\mathrm{AMT}_\mathrm{i}(P_\mathrm{oi})$ calculated following Ref.\ \cite{GC21_HeatReleaze}, should be equally valid for the advanced shell model, i.e.,
$Q_{\rm i}(P_{\rm oi})=Q^\mathrm{AMT}_\mathrm{i}(P_\mathrm{oi})$.
At the same time, one can calculate  $Q_\mathrm{i}(P_\mathrm{inst})$ for a certain CLD+sh model (see Section \ref{Sec_Qdeep}). 
If all heat release in the inner crust is concentrated in its deepest layers,
it is reasonable to assume that 
$Q_\mathrm{i}(P_\mathrm{inst})$
is fully determined by the nuclear physics in the deepest layers of the inner crust (see 
Section\ref{Sec_Qdeep}), and thus $Q_\mathrm{i}(P_\mathrm{inst})$ as calculated for the CLD+sh 
model, should  also be valid for the advanced shell model.

Using the (known) functions 
$Q_\mathrm{i}(P_\mathrm{inst})$ and $Q_\mathrm{i}(P_\mathrm{oi})=Q^\mathrm{AMT}_\mathrm{i}(P_\mathrm{oi})$, it is straightforward to find $P_{\rm inst}$ as a function of $P_{\rm oi}$.
This prediction does not require any additional calculation for the advanced shell model; 
it indicates that the function $P_{\rm inst}(P_{\rm oi})$ does not depend on the form of the shell corrections in the interiors of the inner crust, 
provided that the shell effects are strong enough to prevent the evolution of $Z$ (and thus heat release) up to the deepest layers of the inner crust.

To illustrate this approach, we combined
the function $Q_\mathrm{i}(P_\mathrm{inst})$ calculated in this work 
(Figure \ref{Fig_Q_Pinst}) with the function 
$Q_\mathrm{i}^\mathrm{AMT}(P_\mathrm{oi})$ calculated for 
the BSK24 HFB mass tables \cite{Goriely_ea_Bsk22-26} in Ref.\ \cite{GC21_HeatReleaze} (see inset in Figure \ref{Fig_Pinst_vs_Poi_cor}).
The resulting heuristic prediction for the corrected dependence 
$P_\mathrm{inst}(P_\mathrm{oi})$  is shown in Figure \ref{Fig_Pinst_vs_Poi_cor}.

Comparing Figures \ref{Fig_Pinst_vs_Poi} and \ref{Fig_Pinst_vs_Poi_cor}, one can see that for low $P_\mathrm{oi}$ the corrected predictions for the $Z$-fixed model are closer to the Shell and Shell+Pairing models than in the case of our original calculations in Section \ref{Sec_Res_Shell}. This is because the nuclear physics input for CLD+sh versions of these models differs 
even in the outermost layers of the inner crust.
Specifically, as discussed in Section \ref{Sec_Micr_Shell}, we neglect shell corrections for  the $Z$-fixed model. As a result,  $Q_\mathrm{i}(P_\mathrm{oi})$ calculated within the $Z$-fixed model differs from calculations for Shell and Shell+Pairing models simply because the 
mass of $Z=20$ nuclei predicted by the $Z$-fixed model
differs from the mass of the same nuclei, 
predicted by Shell and Shell+Pairing models.
The corrected results rely on the more accurate nuclear physics in the shallow inner crust region
and should be considered more reliable.

%%%%%%%%%%%%%%%%%%%%%%%%%%%%%%%%%%%
\section{Summary, conclusions and perspectives}\label{Sec_Summary}
%%%%%%%%%%%%%%%%%%%%%%%%%%%%%%%%%%%

For the first time, we present detailed calculations of the nHD crust taking into account proton shell effects in nuclei (see also our preliminary results 
in Ref.\ \cite{GC21_HeatReleaze}). 
Our work clearly demonstrates
that the shell corrections 
have a profound effect on the
FAC models
(compare the results in Sections \ref{Sec_Res_Shell} and \ref{Sec_Res_Smooth}; see also the 
numbered list below).%
%%%%%%%%%
\footnote{
Shell effects also appear to be crucial for calculations made within the traditional approach 
\cite{Fantina_ea18}.}

The general calculation algorithm is described in Section \ref{Sec_Algorithm}; 
it is based on the recently suggested thermodynamic potential $\Psi$ \cite{GKC_psi21}, which should be minimized in the nHD crust. 
Our algorithm can be applied in future studies of nHD accreted crust models.

We numerically construct a family of nHD models, parametrized by the pressure $P_\mathrm{oi}$ at 
the outer-inner crust interface (Section \ref{Sec_Res}).
Next, we analyze this family to determine the range of permissible $P_\mathrm{oi}$ values (see Section \ref{Sec_Poi}).
This two-step procedure is required because the redistribution of unbound neutrons in the inner crust does not allow us to predict $P_\mathrm{oi}$ in advance, as it was in the traditional models, where the outer-inner crust interface was assumed to be associated with the neutron drip point
(e.g., \cite{HZ90,HZ90b,Chamel_etal15_Drip}). 
Within the nHD model, neutrons penetrate to lower pressures, shifting the outer-inner crust interface accordingly.

Obviously, to construct nHD models numerically, one should specify 
a microphysical model. Here, we employ the CLD+sh model of Ref.\ \cite{carreau_ea20_Cryst_CLDsh}, in which the shell effects are added on top of the CLD model based on ETFSI calculations of Refs.\ \cite{Pearson_ea18_bsk22-26,Pearson_ea19_ShellCorr_Errata}. 
The absence of data for odd-$Z$ nuclei in supplementary tables from those 
references forces us to consider two models, labeled as ``Shell'' and ``Shell+Pairing''. 
In the Shell+Pairing model, we add pairing corrections to the energy of odd-$Z$ nuclei using a simplified model, while in the Shell model, these corrections were neglected (see Section \ref{Sec_Micr_Shell} for details). 
As a sensitivity test, we also apply the $Z$-fixed model, in which $Z$ is assumed to be fixed 
at the value $Z=20$
up to the proton drip point (to mimic strong proton shell closure at this value of $Z$).
In addition, we also use the smooth CLD model, where pairing and shell corrections are ignored.

Although numerical details of our calculations 
depend on the applied shell model, 
some universal features can be revealed,
confirming the
preliminary conclusions 
of Ref.\  \cite{GC21_HeatReleaze}:

%%%%%%%%%%
\begin{enumerate}
%%%%%%%%%%%
\item 
Shell corrections suppress beta reactions in most of the inner crust, leading to a composition of 
low-$Z$ ($Z\sim 20$) nuclei rather than the $Z\sim 40$ nuclei predicted by the smooth CLD model 
(see Fig.\ \ref{Fig_EOS_smooth}).

\item 
For all the considered 
microphysical models (smoothed and with shell corrections), 
there exists a minimum pressure $P_\mathrm{oi}^\mathrm{(min)}$, such that for any $P_\mathrm{oi}>P_\mathrm{oi}^\mathrm{(min)}$, there is an instability at pressure $P^\mathrm{inst}(P_\mathrm{oi})$,
leading to the disintegration of nuclei.
Considering the 
function
$\psi(Z)$,
which is the appropriate thermodynamic potential for 
describing 
the nHD inner crust \cite{GKC_psi21}, 
the instability 
reveals itself in
the disappearance of the local minima of $\psi(Z)$  (see Figures \ref{Fig_psi1}, \ref{Fig_PsiProf_smooth}, and \ref{Fig_PsiProf_shell}).
Both the function $P^\mathrm{inst}(P_\mathrm{oi})$ and the pressure $P_\mathrm{oi}^\mathrm{(min)}$ 
are strongly affected by the shell corrections
(see Figure \ref{Fig_Pinst_vs_Poi}).

\item 
$P^\mathrm{inst}$ decreases with an increase in  $P_\mathrm{oi}$ for all considered models 
 (Ref.\ \cite{GC20_DiffEq} demonstrated qualitatively the same behavior for the smooth CLD model 
 based on the SLy4 potential).
$P^\mathrm{inst}$ can be lower than the pressure at the crust-core boundary, $P_\mathrm{cc}$.
If this is the case, the part of the crust between  $P^\mathrm{inst}$ and 
$P_{\rm cc}$ should be considered ``relic'', indicating that it has been formed during the initial phases of accretion from the pristine crust and remained unchanged thereafter
(except for possible small secular evolution associated with the increase in the mass of the 
accreting neutron star).

\item 
$P_\mathrm{oi}$ in the FAC can be constrained using general arguments.

The
lower bound $P_\mathrm{oi}\ge P_\mathrm{oi}^\mathrm{(min)}$ comes from the requirement that the nuclei disintegration mechanism should be active in the FAC to keep the crust
structure stationary and avoid the accumulation of nuclei there.

The upper bound for $P_\mathrm{oi}$ is more complicated (see Section \ref{Sec_Poi}). 
It arises from the condition that for too large $P_\mathrm{oi}$, 
the instability takes place at such a low pressure $P^\mathrm{inst}$ 
that the accreted crust cannot be connected with the stellar core in a thermodynamically consistent way for any composition of the relic part of the crust. 
For the smooth CLD model, this requirement sets  $P_\mathrm{oi}^\mathrm{(min)}$ as an upper bound,
making it
the only allowed value for $P_\mathrm{oi}$. The presence of shell corrections can stabilize the relic crust, allowing for a larger  $P_\mathrm{oi}$, 
which makes
accurate determination of the upper bound challenging. 
By applying our Shell model from Section\ \ref{Sec_Micr_Shell} to the relic part of the crust and assuming 
that the relic region is composed of spherical nuclei, we conclude that 
$P_\mathrm{oi}<P_\mathrm{nd}^{(\mathrm{cat})}$ for pure $^{56}$Fe ash.
However, the validity of this constraint should be checked for the presence of the pasta phases in the relic region.

\item For pure $^{56}$Fe ash, all heat in the inner crust is released in the deep layers, close to the instability point. 
While a similar statement holds true for nHD models based on smooth  CLD \cite{GC21_HeatReleaze}, 
the actual reaction chains are strongly influenced by the shell effects.
The heat release in the deep layers, parameterized by $P^\mathrm{inst}$,  increases as 
$P^\mathrm{inst}$ decreases.

\item There is a minimal heat release $\sim 0.04-0.05$~MeV per accreted nucleon in the deep layers of the inner crust (Section \ref{Sec_MinQ}).
It is not significantly affected by the shell effects.

\item We propose a heuristic approach (Section \ref{Sec_Heurist}), 
enabling the prediction of the properties of the nHD accreted crust for 
advanced nuclear physical models 
that align with atomic mass tables at lower densities 
and resemble CLD+sh 
models at the highest densities.
\end{enumerate}

Subsequent studies should analyze to what extent our results are sensitive to the ash composition (the present work is limited to the pure $^{56}$Fe ash), nuclear physical models, and so on. 
These problems can be 
approached using the algorithm suggested and applied here (see Section \ref{Sec_Algorithm}). 
We expect that the majority of our qualitative conclusions will remain unaffected 
even in this, more general situation.

%%%%%%%%%%%%%%%%%%%%%%%%%%%%%%%%%%%%%%%%%%%%%%%%%%%%%%%%%
\begin{acknowledgments}
%%%%%%%%%%%%%%%%%%%%%%%%%%%%%%%%%%%%%%%%%%%%%%%%%%%%%%%%%
This work was 
supported by Russian Science Foundation (grant No. 22-12-00048, 
\href{https://rscf.ru/project/22-12-00048/}{https://rscf.ru/project/22-12-00048/}).
In the final stages of the work, one of the authors (MEG) was on a long-term visit at the Weizmann Institute of Science (WIS).
MEG acknowledges the support of the visit by the Simons Foundation and WIS. MEG is also grateful to the Department of Particle Physics \& Astrophysics at WIS for their hospitality and excellent working conditions. 

\end{acknowledgments}
%%%%%%%%%%%%%%%%%%%%%%%%%%%%%%%%%%%%%%%%%%%%%%%%%%%%%%%%%

%%%%%%%%%%%%%%%%%%%%%%%%%%%%%%%%%%%%%%%%%%%%%%%%%%%%%%%%%%%%%%%%%%%%%%%%%%%%%%%%%%%%%%
%\bibliography{literature}

\end{document}